\documentclass[apj]{emulateapj}

\usepackage{mathptmx}
\usepackage{verbatim}
\usepackage{graphicx}
\usepackage{epstopdf}
\usepackage{amsmath}
\usepackage{textcomp}
\usepackage{latexsym}
\usepackage{multirow}
\usepackage{wasysym}

\newcommand{\ltsim}{\raisebox{-.5ex}{$\;\stackrel{<}{\sim}\;$}}
\newcommand{\gtsim}{\raisebox{-.5ex}{$\;\stackrel{>}{\sim}\;$}}
\newcommand{\kms}{\ifmmode {\rm km\ s}^{-1} \else km s$^{-1}$\fi}
\newcommand{\lledd}{$L/L_{\rm Edd}$}
\newcommand{\msun}{$M_{\odot}$}
\newcommand{\mbh}{$M_{\rm BH}$}
\newcommand{\et}{et al.\ }
\newcommand{\xray}{\hbox{X-ray}}
\newcommand{\aox}{$\alpha_{\rm ox}$}
\newcommand{\daox}{$\Delta\alpha_{\rm ox}$}
\newcommand{\nh}{$N_{\rm H}$}
\newcommand{\xmm}{{\sl XMM-Newton}}
\newcommand{\chandra}{{\sl Chandra}}
\newcommand{\swift}{{\sl Swift}}
\newcommand{\hb}{H$\beta$}

\journalinfo{The Astrophysical Journal, ???:??? (??pp), 2014 ??? ??}
\slugcomment{Received 2013 December 1; accepted 2014 January 17; published 2014 ??? ??}

\shortauthors{SHEMMER ET AL.}
\shorttitle{X-RAY MONITORING OF HIGH-REDSHIFT QUASARS}

\begin{document}
\title{Exploratory X-ray Monitoring of Luminous Radio-Quiet Quasars at High Redshift: \\ Initial Results}

\author{
Ohad~Shemmer,\altaffilmark{1}
W.~N.~Brandt,\altaffilmark{2,3}
Maurizio~Paolillo,\altaffilmark{4,5}
Shai~Kaspi,\altaffilmark{6,7}
Cristian~Vignali,\altaffilmark{8}
Matthew~S.~Stein,\altaffilmark{1}
Paulina~Lira,\altaffilmark{9}
Donald~P.~Schneider,\altaffilmark{2,3}
and Robert~R.~Gibson\altaffilmark{10}
}
\altaffiltext{1}
		   {Department of Physics, University of North Texas, Denton, TX 76203, USA; ohad@unt.edu}
\altaffiltext{2}
		   {Department of Astronomy \& Astrophysics, The Pennsylvania State University, University Park, PA 16802, USA}
\altaffiltext{3}
                     {Institute for Gravitation and the Cosmos, The Pennsylvania State University, University Park, PA 16802, USA}
\altaffiltext{4}
                     {Dipartimento di Scienze Fisiche, Universit\`{a} Federico II di Napoli, via Cinthia 6, I-80126 Napoli, Italy}
\altaffiltext{5}
                     {ASI Science Data Center, Via del Politecnico snc, 00133 Rome, Italy}
\altaffiltext{6}
		   {School of Physics \& Astronomy and the Wise Observatory, Tel Aviv University, Tel Aviv 69978, Israel}
\altaffiltext{7}
		   {Department of Physics, Technion, Haifa 32000, Israel}
\altaffiltext{8}
		  {Dipartimento di Astronomia, Universit\`{a} degli studi di Bologna, via Ranzani 1, I-40127 Bologna, Italy}
\altaffiltext{9}
 		   {Departamento de Astronomia, Universidad de Chile, Camino del Observatorio 1515, Santiago, Chile}
\altaffiltext{10}
                      {Department of Astronomy, University of Washington, Box 351580, Seattle, WA 98195, USA}

\begin{abstract}
We present initial results from an exploratory \xray\ monitoring project of two groups of comparably luminous radio-quiet quasars (RQQs). The first consists of four sources at \hbox{$4.10\leq z\leq 4.35$}, monitored by \chandra, and the second is a comparison sample of three sources at
\hbox{$1.33\leq z\leq 2.74$}, monitored by \swift. Together with archival \xray\ data, the total rest-frame temporal baseline spans \hbox{$\sim2-4$}~yr and \hbox{$\sim5-13$}~yr for the first and second group, respectively. Six of these sources show significant \xray\ variability over rest-frame timescales of \hbox{$\sim10^2 - 10^3$}~d; three of these also show significant \xray\ variability on rest-frame timescales of \hbox{$\sim1-10$}~d. The
\xray\ variability properties of our variable sources are similar to those exhibited by nearby and far less luminous active galactic nuclei (AGNs). While
we do not directly detect a trend of increasing \xray\ variability with redshift, we do confirm previous reports of luminous AGNs exhibiting \xray\
variability above that expected from their luminosities, based on simplistic extrapolation from lower luminosity sources. This result may be attributed to luminous sources at the highest redshifts having relatively high accretion rates. Complementary UV-optical monitoring of our sources shows that variations in their optical-\xray\ spectral energy distribution are dominated by the \xray\ variations. We confirm previous reports of \xray\ spectral variations in one of our sources, HS~1700$+$6416, but do not detect such variations in any of our other sources in spite of \xray\ flux variations of up to a factor of $\sim4$. This project is designed to provide a basic assessment of the \xray\ variability properties of RQQs at the highest accessible redshifts that will serve as a benchmark for more systematic monitoring of such sources with future \xray\ missions.
\end{abstract}


\section{INTRODUCTION}
\label{sec:introduction}

Active galactic nuclei (AGNs) exhibit intensity fluctuations across the electromagnetic spectrum on timescales ranging from minutes to decades (e.g., Ulrich \et 1997). While the source of these fluctuations is not yet fully understood, the bulk of the observed AGN variability is thought to arise in the accretion disk, corona, broad emission line region, and jet. Albeit its non-periodic nature, AGN variability provides valuable information on the size and structure of the innermost regions of the central engine (e.g., Collier \et 1998; Shemmer \et 2001; Chartas \et 2012). In particular, it can be used to estimate the masses of supermassive black holes (BHs; e.g., Kaspi \et 2000a; Lu \& Yu 2001; Papadakis 2004; O'Neill \et 2005; McHardy \et 2006; Bentz \et 2009; Ponti \et 2012). Light-crossing time considerations imply that the timescales (amplitudes) of AGN flux variations should increase (decrease) with the emission-region size. Continuum variations in the X-rays, the bulk of which are thought to be emitted from the inner $\sim10$ gravitational radii, are thus faster and stronger relative to those in, e.g., the optical band, and therefore provide a more efficient way of probing the central engines of AGNs, at least on relatively short timescales.

\begin{deluxetable*}{lcccccccc}
\tablecaption{Basic \xray, Optical, and Radio Properties of the \chandra\ Sources}
\tablecolumns{9}
\tablehead{
\colhead{} &
\colhead{} &
\colhead{} &
\colhead{} &
\colhead{Galactic \nh\tablenotemark{a}} &
\colhead{$\log \nu L_{\nu} (2$\,keV)\tablenotemark{b}} &
\colhead{$\log \nu L_{\nu} (2500$\,\AA)\tablenotemark{b}} &
\colhead{} &
\colhead{} \\
\colhead{Quasar} &
\colhead{$\alpha$ (J2000.0)} &
\colhead{$\delta$ (J2000.0)} &
\colhead{$z$} &
\colhead{(10$^{20}$\,cm$^{-2}$)} &
\colhead{(erg s$^{-1}$)} &
\colhead{(erg s$^{-1}$)} &
\colhead{\aox\tablenotemark{c}} &
\colhead{$R\tablenotemark{d}$}
}
\startdata
Q~0000$-$263         & 00 03 22.9 & $-$26 03 16.8 & 4.10 & 1.67 & 45.5 & 47.4 & $-1.70$ & $<4.7$ \\
BR~0351$-$1034    & 03 53 46.9 &  $-$10 25 19.0 & 4.35 & 4.08 & 45.2 & 46.9 & $-1.69$ & $1.2$ \\
PSS~0926$+$3055 & 09 26 36.3 & $+$30 55 05.0 & 4.19 & 1.89 & 45.7 & 47.7 & $-1.76$ & $<0.4$ \\
PSS~1326$+$0743 & 13 26 11.9 & $+$07 43 58.4 & 4.17 & 2.01 & 45.5 & 47.5 & $-1.76$ & $<0.7$
\enddata
\tablenotetext{a}{Obtained using the \nh\ tool at http://heasarc.gsfc.nasa.gov/cgi-bin/Tools/w3nh/w3nh.pl}
\tablenotetext{b}{Obtained from Table~3 of Shemmer \et (2005), assuming a hard-\xray\ photon index of $\Gamma = 2.0$ (Shemmer \et 2005) and a UV continuum of the form $f_{\nu} \propto \nu^{-0.5}$ (Vanden Berk \et 2001).}
\tablenotetext{c}{The \xray-UV power-law slope, \aox, is defined as \aox$\displaystyle =\frac{\log(f_{2\,\rm keV}/f_{\rm 2500\,\mbox{\scriptsize\AA}})}{\log(\nu_{2\,\rm keV}/\nu_{\rm 2500\,\mbox{\scriptsize\AA}})}$, where $f_{2\,\rm keV}$ and $f_{\rm 2500\,\mbox{\scriptsize\AA}}$ are the flux densities at rest-frame 2~keV and 2500~\AA, respectively. The \aox\ values are obtained from the $\log \nu L_{\nu} (2$\,keV) and
$\log \nu L_{\nu} (2500$\,\AA) data.}
\tablenotetext{d}{Radio-loudness parameter (Kellermann \et 1989). The radio flux for BR~0351$-$1034 was obtained from Isaak \et (1994); upper limits on the radio flux for PSS~0926$+$3055 and PSS~1326$+$0743 were obtained from Becker \et (1995), and from Condon \et (1998) for Q~0000$-$263.}
\label{tab:chandra_props}
\end{deluxetable*}

While \xray\ variability in nearby AGNs has been the subject of intensive study (e.g., Nandra \et 1997; Fiore \et 1998; Turner \et 1999; Uttley \et 2002; Markowitz \et 2003), luminous quasars, found mostly at $z>1$ during the main growth phase of their supermassive BHs, have been relatively neglected until recently (see, e.g., Vagnetti \et 2011; Gibson \& Brandt 2012). Being typically $\approx10^3-10^4$ times more luminous than the highly variable local Seyfert galaxies, the emission regions of luminous quasars are expected to be physically larger and thus have slower and milder variations (e.g., Barr \& Mushotzky 1986; Green \et 1993; Lawrence \& Papadakis 1993). Surprisingly, however, the past $\approx15$~yr has seen tentative evidence suggesting evolution in the AGN \xray\ variability properties, where sources of matched luminosity appear more \xray\ variable as redshift increases up to $z \sim 4$ (Almaini \et 2000; Manners \et 2002; Paolillo \et 2004, hereafter P04). Almaini \et (2000) found that $z>0.5$ sources in their {\sl ROSAT} deep, flux-limited quasar sample did not show an anticorrelation between variability amplitude and luminosity as observed in local AGN. Using a large {\sl ROSAT} sample of bright radio-quiet quasars (RQQs\footnote{Throughout this work we define radio-quiet AGNs as sources having $R = f_{\nu} (5~{\rm GHz}) / f_{\nu} (4400~{\rm \AA}) < 10$, following Kellermann \et (1989). We also assume radio and optical-UV continua of the form $f_{\nu} \propto \nu^{-0.5}$ for computing $R$.}) at $z\sim 2-4$, Manners \et (2002) reported significant \xray\ variations on rest-frame timescales as short as $\approx1$~d. Utilizing the 1~Ms \chandra\ Deep Field-South (CDF-S) survey, P04 found that at $2\ltsim z \ltsim 3$ AGNs become more \xray\ variable on rest-frame timescales of \hbox{$\sim1-100$}~d, and their variability amplitudes are larger than expected from extrapolations of their low-redshift counterparts. Significant \xray\ flux variations between two epochs separated by rest-frame timescales of $\sim10-100$~d for sources at $z\sim 4-6$ were also reported by Shemmer \et (2005). Collectively, these pieces of evidence suggest evolution of the \xray\ variability mechanism, the \xray\ emitting region size, or the accretion rate (e.g., Manners \et 2002; P04).

However, evolutionary scenarios for these phenomena are perhaps puzzling given that the basic \xray\ spectral properties of optically-selected AGN (i.e., hard-\xray\ power-law photon index, intrinsic absorption, and the optical-\xray\ spectral slope) have not evolved significantly over cosmic time up to $z\sim6$ (e.g., Shemmer \et 2005, 2006a; Vignali \et 2005; Steffen \et 2006; Just \et 2007), consistent with findings from longer wavelengths (but see also Bechtold \et 2003; Grupe \et 2006; Kelly \et 2007). \xray\ monitoring of large numbers of high-redshift quasars is therefore crucial for resolving this possible discrepancy and for testing evolutionary scenarios of the AGN central engine. It is particularly important to monitor quasars at $z>4$ regularly to add leverage at the highest accessible redshifts and to probe sources at the earliest stages of their formation. In spite of the fact that the number of \xray-detected quasars at $z>4$ has grown more than tenfold since the 12 such sources known about a decade ago (Kaspi \et 2000b), until recently only two \xray\ epochs were available for a handful of such quasars (e.g., Shemmer \et 2005; Gibson \& Brandt 2012) and, prior to this work, no systematic \xray\ monitoring of such sources has been performed.

Quasar monitoring at the highest redshifts is challenging due to cosmic time dilation and the low flux levels of the sources involved (see, e.g., Kaspi \et 2007). \xray\ monitoring of such sources is particularly challenging since observations at higher redshifts necessarily involve harder X-rays, which are more difficult to detect given the steep quasar spectral energy distribution (SED) in the accessible energy range (Haardt \& Maraschi 1991 and references therein). Moreover, observations of such sources are currently feasible with only two observatories, \chandra\ \xray\ Observatory (hereafter \chandra; Weisskopf \et 2000) and \xmm\ (Jansen \et 2001). Since the relatively faster variations in the \xray\ band compared to the optical-UV provide some compensation against cosmic time dilation, effective monitoring campaigns of high-redshift quasars may be performed within the typical lifetime of an \xray\ observatory. We therefore embarked on an exploratory \xray\ monitoring project in order to extend our knowledge of quasar \xray\ variability to the highest accessible redshifts while the current flagship \xray\ observatories are operational.

This paper presents our monitoring strategy and the initial results of this project. We describe our target selection, the observations and their processing in \S~\ref{sec:observations}, and present our basic findings in \S~\ref{sec:results}. Our results are discussed in \S~\ref{sec:discussion} and briefly summarized in \S~\ref{sec:conclusions}. Luminosity distances were computed using the standard cosmological model
(\hbox{$\Omega_{\Lambda}=0.7$}, \hbox{$\Omega_{\rm M}=0.3$}, and \hbox{$H_0=70$~\kms~Mpc$^{-1}$}; e.g., Spergel \et 2007), and Galactic column densities were obtained from Dickey \& Lockman (1990).

\begin{deluxetable*}{lcccccccc}
\tablecolumns{9}
\tablecaption{Basic \xray, Optical, and Radio Properties of the \swift\ Sources}
\tablehead{
\colhead{} &
\colhead{} &
\colhead{} &
\colhead{} &
\colhead{Galactic \nh\tablenotemark{a}} &
\colhead{$\log \nu L_{\nu} (2$\,keV)\tablenotemark{b}} &
\colhead{$\log \nu L_{\nu} (2500$\,\AA)} &
\colhead{} &
\colhead{}  \\
\colhead{Quasar} &
\colhead{$\alpha$ (J2000.0)} &
\colhead{$\delta$ (J2000.0)} &
\colhead{$z$} &
\colhead{(10$^{20}$\,cm$^{-2}$)} &
\colhead{(erg s$^{-1}$)} &
\colhead{(erg s$^{-1}$)} &
\colhead{\aox} &
\colhead{$R\tablenotemark{e}$}
}
\startdata
PG~1247$+$267   & 12 50 05.724 & $+$26 31 07.55 & 2.04 &  0.90 & 45.7 & 47.4\tablenotemark{c} & $-1.69$ & $0.4$ \\
PG~1634$+$706   & 16 34 29.000 & $+$70 31 32.40 & 1.33 &  4.48 & 45.7 & 47.5\tablenotemark{c} & $-1.69$ & $0.4$ \\
HS~1700$+$6416 & 17 01 00.620 & $+$64 12 09.12 & 2.74 & 2.66 & 45.4 & 47.6\tablenotemark{d} & $-1.83$ & $<0.1$
\enddata
\tablenotetext{a}{Obtained using the \nh\ tool at http://heasarc.gsfc.nasa.gov/cgi-bin/Tools/w3nh/w3nh.pl}.
\tablenotetext{b}{Obtained from the mean \xray\ flux of the \swift\ observations, assuming a hard-\xray\ photon index of $\Gamma = 2.0$
(see \S~\ref{sec:new_swift}).}
\tablenotetext{c}{Obtained from the mean optical flux of the \swift\ observations (see \S~\ref{sec:uvot}); consistent with the average optical flux of the source from Trevese \et (2007).}
\tablenotetext{d}{Obtained from  the mean optical flux of the \swift\ observations (see \S~\ref{sec:uvot}); consistent with the average optical flux of the source from Kaspi \et (2007).}
\tablenotetext{e}{Radio-loudness parameter (Kellermann \et 1989). Radio fluxes for PG~1247$+$267 and PG~1634$+$706 were obtained from Kellermann \et (1989); an upper limit on the radio flux of HS~1700$+$6416 was obtained from Becker \et (1995).}
\label{tab:swift_props}
\end{deluxetable*}

\section{TARGET SELECTION, OBSERVATIONS, AND DATA REDUCTION}
\label{sec:observations}

\subsection{Strategy and Target Selection}
\label{sec:targets}

The aim of this exploratory monitoring project is to obtain, through time-series analyses, a qualitative assessment of the basic \xray\ variability properties, such as amplitudes and timescales, for a small and carefully-selected sample of optically-selected RQQs at high redshift and compare them with the \xray\ variability properties of AGNs at lower luminosities and redshifts. The selection of RQQs was intended to minimize
interference of jet-related \xray\ variations with those of the accretion-disk-corona system. The results of this project will guide more ambitious systematic \xray\ observations of larger numbers of high-redshift sources with future \xray\ missions. The ultimate goals are to identify the parameters that drive AGN \xray\ variability and to test whether \xray\ variability has evolved with cosmic time.

For our sample of RQQs at the highest accessible redshifts, we selected four sources at \hbox{$4.10\leq z\leq 4.35$} from Shemmer \et (2005) that had at least two archival \xray\ observations, per source, and were bright enough for economical \chandra\ observations (hereafter, the ``\chandra\ sources''). These sources include Q~0000$-$263, BR~0351$-$1034, PSS~0926$+$3055, and PSS~1326$+$0743, all of which are optically-selected quasars; their basic \xray, optical, and radio properties appear in Table~\ref{tab:chandra_props}. Two of these sources, Q~0000$-$263 and BR~0351$-$1034, were discovered as part of the objective prism survey and the color survey with the Automatic Plate Measuring machine, respectively, using the UK Schmidt telescope in an effort to search for $z>4$ quasars (e.g., Webb \et 1988; Irwin \et 1991). The other two sources, PSS~0926$+$3055 and PSS~1326$+$0743, were discovered as $z>4$ quasars using multicolor selection with the Palomar Digital Sky Survey (e.g., Djorgovski \et 1998). Beginning with Cycle~12, these sources are observed once per \chandra\ Cycle. The first \xray\ epoch for each of these sources was obtained by shallow {\sl ROSAT} (Truemper 1982) or \chandra\ observations. The second epoch for each of these was obtained by a lengthy \xmm\ observation which also enabled searching for rapid \xray\ variations on timescales of $\sim1$~hr (see Shemmer \et 2005 for more details about the first two epochs).

\begin{figure}
\epsscale{1.2}
\plotone{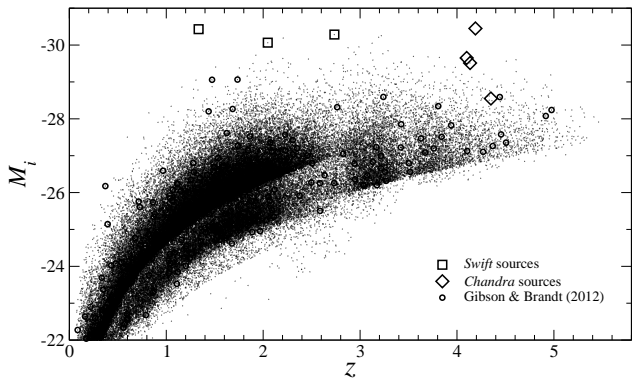}
\caption{Absolute $i$-band magnitude versus redshift for $105,783$ SDSS quasars (dots; Schneider \et 2010). The \chandra\ (\swift) sources are marked with diamonds (squares). SDSS quasars from the Gibson \& Brandt (2012) study are marked with circles. One \chandra\ source,
PSS~1326$+$0743, and two \swift\ sources, PG~1247$+$267 and HS~1700$+$6416, are also SDSS quasars.}
\label{fig:L_z}
\end{figure}

In order to distinguish between intrinsic and potential environmental effects on quasar variability it is necessary to break the strong luminosity-redshift dependence inherent in most quasar surveys. We therefore complemented the \chandra\ sources with a comparison sample of three RQQs with luminosities that are similar to those of the \chandra\ sample, but at considerably lower redshifts ($1.33\leq z \leq 2.74$). The sources of the comparison sample are sufficiently \xray\ bright for monitoring with the \swift\ Gamma-ray Burst Explorer (Gehrels \et 2004; hereafter, the ``\swift\ sources''). The \chandra\ sources, on the other hand, are too faint for \swift. The \swift\ sources were selected to have the largest number of archival \xray\ epochs without compromising the luminosity and redshift requirements. These sources include the optically-selected quasars,
PG~1247$+$267, PG~1634$+$706, and HS~1700$+$6416; their basic \xray, optical, and radio properties appear in Table~\ref{tab:swift_props}.
The first two of these were discovered as part of the Bright Quasar Survey designed to identify UV-excess sources (e.g., Schmidt \& Green 1983).
The third quasar, HS~1700$+$6416, was discovered in the course of the Hamburg wide-angle objective prism survey for bright quasars on the northern sky (Reimers \et 1989).

The selection process of both samples, inevitably, introduces additional cross-calibration uncertainties stemming from the inclusion of datasets from different \xray\ missions; these uncertainties are discussed further below. Nevertheless, our strategy maximally utilizes the \xray\ archive and allows for the most efficient exploratory \xray\ monitoring campaign of high-redshift RQQs. Figure~\ref{fig:L_z} shows the luminosities and redshifts of the sources in our two samples against the Sloan Digital Sky Survey (SDSS; York \et 2000) quasar catalog of Schneider \et (2010). Our two samples clearly mark the high-luminosity envelope of optically-selected quasars and thereby probe \xray\ variability in a completely new region of the $L-z$ parameter space. Also shown in Fig.~\ref{fig:L_z} are the 264 SDSS quasars\footnote{Two of these sources, SDSS~J083454.89$+$553421.1 and SDSS~J150407.51$-$024816.5, are AGNs which appear in an earlier version of the SDSS quasar catalog (Schneider \et 2007), and were dropped from the Schneider \et (2010) catalog since they did not meet the $M_i <-23$ criterion that would qualify them as quasars due to improved SDSS photometry.} from Gibson \& Brandt (2012). Eight of their sources are at $z>4$, but none has more than two \xray\ epochs. Our strategy is therefore complementary to the one used by Gibson \& Brandt (2012), and to the survey-based studies of, e.g., P04, Mateos \et (2007), Vagnetti \et (2011, 2013), and Ponti \et (2012). Finally, both the \chandra\ and \swift\ sources are representative of highly luminous type~1 (i.e., unobscured) RQQs (with radio-loudness values almost an order of magnitude below the traditional $R=10$ cutoff). They all display typical blue continua with no broad absorption lines and typical emission-line properties (e.g., Green \et 1980; Neugebauer \et 1987; Baldwin \et 1989; Schneider \et 1989; Storrie-Lombardi \et 1996; Vignali \et 2003; Schneider \et 2010), hard-\xray\ power-law photon indices in the range $\Gamma \sim 1.8-2.2$ (Page \et 2004; Shemmer \et 2005; Lanzuisi \et 2012), and optical-\xray\ spectral slopes (hereafter \aox; see Table~\ref{tab:chandra_props}) consistent within
$1~\sigma$ with expectations from their UV luminosities (Steffen \et 2006; Just \et 2007).

\subsection{New \chandra\ Observations}
\label{sec:new_chandra}

We obtained \chandra\ snapshot observations of the \chandra\ sources in Cycles~12~and~13 \hbox{(2011--2012)}; the observation log appears in
Table~\ref{tab:chandra_log}. These data were obtained with the Advanced CCD Imaging Spectrometer (ACIS; Garmire \et 2003) with the S3 CCD at the aimpoint using Faint mode for the event telemetry format in all the observations. Only events with grades of 0,~2,~3,~4,~and~6 were considered
in the analysis, which was performed using standard \chandra\ Interactive Analysis of Observations
({\sc ciao)\footnote{http://cxc.cfa.harvard.edu/ciao/} v4.1} routines. No background flares are present in these observations. The \xray\ counts
in the observed-frame ultrasoft band (\hbox{0.3--0.5~keV}), soft band (\hbox{0.5--2~keV}), hard band (\hbox{2--8~keV}), and full band
(\hbox{0.5--8~keV}) were extracted with the {\sc wavdetect} thread (Freeman \et 2002) using wavelet transforms (with wavelet scale sizes of 1,~1.4,~2,~2.8,~and~4 pixels) and a false-positive probability threshold of 10$^{-3}$; these \xray\ counts are reported in
Table~\ref{tab:chandra_counts}. The use of a relatively large false-positive probability threshold (10$^{-3}$) is justified mainly due to the accurate a priori source positions and, indeed, in most of the bands, source detection was also achieved with a more conservative false-positive probability threshold of 10$^{-6}$. A manual inspection of all the \chandra\ images yielded consistent results with the {\sc wavdetect} photometry. For each source, Table~\ref{tab:chandra_counts} also lists the band ratio, which is the ratio between the counts in the hard band and the soft band, the effective power-law photon index,\footnote{The effective power-law photon index $\Gamma$, defined as $N(E)\propto E^{-\Gamma}$, was derived from the band ratio using the \chandra\ {\sc pimms} v4.4 tool at http://cxc.harvard.edu/toolkit/pimms.jsp} the soft-band count rate, and the flux density at rest-frame 2~keV. Galactic absorption-corrected fluxes in the observed-frame \hbox{0.5--2~keV} band were obtained using the
\chandra\ {\sc pimms} v4.4 tool, assuming a power-law model with $\Gamma=2.0$, and are discussed in \S~\ref{sec:results_amplitudes}.

\subsection{New \swift\ Observations}
\label{sec:new_swift}

We obtained \swift\ snapshot observations of the \swift\ sources in Cycles 3, 4, and 8 \hbox{(2007--2013)}; the observation log appears in
Table~\ref{tab:swift_log}. The observations were obtained with the \xray\ Telescope ({\sl XRT}; Burrows \et 2005) simultaneously with the Ultraviolet Optical Telescope ({\sl UVOT}; Roming \et 2005) on board \swift. The {\sl UVOT} filters used in each observation are given in
Table~\ref{tab:swift_log}; the analysis of the {\sl UVOT} data is discussed in \S~\ref{sec:uvot}. The {\sl XRT} Level~2 data were analyzed using standard {\sc xselect}\footnote{http://heasarc.gsfc.nasa.gov/docs/software/lheasoft/ftools/xselect/index.html} routines from {\sc ftools} (Blackburn 1995). For each source, counts in the observed-frame $0.2-10$~keV band were extracted from source regions with an aperture having a radius of 20 pixels (corresponding to 47.2\arcsec) around the quasars' optical positions (this includes 90\% of the energy of the point spread function at an observed-frame energy of 1.5~keV). Background counts were extracted from source-free regions that were six times larger (in area) than the source regions. The net counts are listed in Table~\ref{tab:swift_log}. The corresponding \xray\ fluxes were obtained using
WebPIMMS\footnote{http://heasarc.nasa.gov/Tools/w3pimms.html}, corrected for Galactic absorption and assuming a power-law photon index of
$\Gamma = 2.0$, and are discussed in \S~\ref{sec:results_amplitudes}.

Ideally, we would have performed the \xray\ time-series analyses in the same rest-frame energy band for all our sources. For the \chandra\ sources, we adopted the observed-frame \hbox{$0.5 - 2.0$}~keV energy band, in which we obtain the largest fraction of the total counts from each source. 
Since the redshift of each \chandra\ source is \hbox{$z\sim4.2$}, this band corresponds to the rest-frame \hbox{$\sim2.6-10.4$}~keV energy band in each of these sources. We therefore repeated the \swift\ data analysis as described above but filtered the {\sl XRT} event files to include only the rest-frame \hbox{$\sim2.6-10.4$}~keV energy range for each \swift\ source. This filtering effectively reduced the number of counts by a factor of $\sim2$ and reduced the signal-to-noise (S/N) level by $\sim50$\% for each \swift\ observation. Using a Spearman-rank correlation we find that the \xray\ fluxes obtained from the filtered observations are correlated with the corresponding fluxes in the observed-frame \hbox{$0.2 - 10$}~keV band at
$>99.9$\% confidence level. This result is consistent with the results of P04 who did not find significant changes in their time-series analysis when considering a narrower observed-frame energy band that corresponded to the rest-frame energy band of most of their sources.
We therefore perform time-series analyses using the observed-frame \hbox{$0.2 - 10$}~keV band for the \swift\ sources, thus maximizing the S/N levels of their \swift\ observations.

\begin{deluxetable}{lclcc}
\tablecolumns{5}
\tablecaption{Log of New \chandra\ Observations of the \chandra\ Sources}
\tablehead{
\colhead{} &
\colhead{} &
\colhead{} &
\colhead{} &
\colhead{Exp. Time\tablenotemark{a}} \\
\colhead{Quasar} &
\colhead{Cycle} &
\colhead{Obs. Date} & 
\colhead{Obs. ID} &
\colhead{(ks)}
}
\startdata
Q~0000$-$263         & 12 & 2011 Aug 29 & \dataset [ADS/Sa.CXO#obs/12795] {12795} & 9.92 \\
                                     & 13 & 2012 Sep 2   & \dataset [ADS/Sa.CXO#obs/14215] {14215} & 9.93 \\
BR~0351$-$1034    & 12 & 2011 Sep 23 & \dataset [ADS/Sa.CXO#obs/12796] {12796} & 9.92 \\
                                     & 13 & 2011 Oct 28  & \dataset [ADS/Sa.CXO#obs/14218] {14218} & 9.84 \\
PSS~0926$+$3055 & 12 & 2011 Mar 3   & \dataset [ADS/Sa.CXO#obs/12793] {12793} & 4.98 \\
                                     & 13 & 2012 Jan 13 & \dataset [ADS/Sa.CXO#obs/14209] {14209} & 4.98 \\
PSS~1326$+$0743 & 12 & 2011 Mar 7   & \dataset [ADS/Sa.CXO#obs/12794] {12794} & 4.98 \\
                                     & 13 & 2012 Apr 30 & \dataset [ADS/Sa.CXO#obs/14212] {14212} & 5.00
\enddata
\tablenotetext{a}{The \chandra\ exposure time has been corrected for detector dead time.}
\label{tab:chandra_log}
\end{deluxetable}

\begin{deluxetable*}{lccccclllc}
\tablecolumns{10}
\tablecaption{Basic \xray\ Measurements from New \chandra\ Observations of the \chandra\ Sources}
\tablehead{ 
\colhead{} &
\colhead{} &
\multicolumn{4}{c}{Counts$^{\rm a}$} \\
\cline{3-6} \\
\colhead{Quasar} &
\colhead{Cycle} &
\colhead{0.3--0.5~keV} &
\colhead{0.5--2~keV} & 
\colhead{2--8~keV} &
\colhead{0.5--8~keV} & 
\colhead{Band Ratio\tablenotemark{b}} &
\colhead{$\Gamma$\tablenotemark{b}} &
\colhead{Count Rate\tablenotemark{c}} &
\colhead{$f_{2~\rm keV}$\tablenotemark{d}}
}
\startdata
Q~0000$-$263 & 12 & {\phn}4.0$^{+3.2}_{-1.9}$ & {\phn}54.3$^{+8.4}_{-7.3}$ & {\phn}14.8$^{+4.9}_{-3.8}$ & {\phn}69.0$^{+9.4}_{-8.3}$ &
{\phn}0.27$^{+0.10}_{-0.08}$ & {\phn}1.9$\pm$0.3 & {\phn}5.47$^{+0.85}_{-0.74}$ & 17.22 \\
& 13 & {\phn}4.9$^{+3.4}_{-2.1}$ & {\phn}44.7$^{+7.7}_{-6.7}$ & {\phn}18.6$^{+5.4}_{-4.3}$ & {\phn}63.3$^{+9.0}_{-7.9}$ &
{\phn}0.42$^{+0.14}_{-0.11}$ & {\phn}1.5$\pm$0.3 & {\phn}4.50$^{+0.78}_{-0.67}$ & 14.24 \\
BR~0351$-$1034 & 12 & $<3.0$ & {\phn}11.8$^{+4.5}_{-3.4}$ & $2.9^{+2.9}_{-1.6}$ & {\phn}14.7$^{+4.9}_{-3.8}$ & {\phn}$0.24^{+0.26}_{-0.15}$ & {\phn}$2.1^{+0.9}_{-0.7}$ & {\phn}1.19$^{+0.46}_{-0.34}$ & 4.20 \\
& 13 & $<3.0$ & {\phn}9.8$^{+4.3}_{-3.1}$ & $2.9^{+2.9}_{-1.6}$ & {\phn}$12.7^{+4.7}_{-3.5}$ & {\phn}$0.29^{+0.32}_{-0.18}$ &
{\phn}$1.9^{+0.9}_{-0.7}$ & {\phn}1.00$^{+0.43}_{-0.31}$ & 3.52 \\
PSS~0926$+$3055 & 12 & {\phn}2.0$^{+2.7}_{-1.3}$ & {\phn}32.6$^{+6.8}_{-5.7}$ & {\phn}10.8$^{+4.4}_{-3.2}$ & {\phn}44.3$^{+7.7}_{-6.6}$ & {\phn}0.33$^{+0.15}_{-0.11}$  & {\phn}1.7$^{+0.4}_{-0.3}$ & {\phn}6.53$^{+1.36}_{-1.14}$ & 21.05 \\
& 13 & $<4.8$ & {\phn}22.7$^{+5.8}_{-4.7}$ & {\phn}8.0$^{+4.0}_{-2.8}$ & {\phn}30.7$^{+6.6}_{-5.5}$ & {\phn}0.35$^{+0.20}_{-0.14}$  &
{\phn}1.7$^{+0.5}_{-0.4}$ & {\phn}4.57$^{+1.18}_{-0.95}$ & 14.90 \\
PSS~1326$+$0743 & 12 & {\phn}3.0$^{+2.9}_{-1.6}$ & {\phn}33.7$^{+6.9}_{-5.8}$ & {\phn}9.9$^{+4.3}_{-3.1}$ & {\phn}43.5$^{+7.7}_{-6.6}$ & {\phn}0.29$^{+0.14}_{-0.10}$  & {\phn}1.9$\pm$0.4 & {\phn}6.75$^{+1.34}_{-1.16}$ & 21.75 \\
& 13 & $2.0^{+2.7}_{-1.3}$ & {\phn}32.4$^{+6.8}_{-5.7}$ & {\phn}11.9$^{+4.6}_{-3.4}$ & {\phn}44.4$^{+7.7}_{-6.6}$ & {\phn}0.37$^{+0.16}_{-0.12}$  &
{\phn}1.7$^{+0.4}_{-0.3}$ & {\phn}6.49$^{+1.35}_{-1.13}$ & 21.00
\enddata
\tablenotetext{a}{Errors on the \xray\ counts were computed according to Tables~1 and 2 of Gehrels (1986) and correspond to the 1$\sigma$ level; these were calculated using Poisson statistics. The upper limits are at the 95\% confidence level and were computed according to Kraft \et (1991). Upper limits of 3.0 and 4.8 indicate that 0 and 1 \xray\ counts, respectively, have been found within an extraction region of radius 1\arcsec\ centered on the optical position of the quasar (considering the background within this source-extraction region to be negligible).}
\tablenotetext{b}{We calculated errors at the 1$\sigma$ level for the band ratio (the ratio between the \hbox{2--8~keV} and \hbox{0.5--2~keV} counts) and effective photon index following the ``numerical method'' described in \S~1.7.3 of Lyons (1991); this avoids the failure of the standard approximate-variance formula when the number of counts is small (see \S~2.4.5 of Eadie \et 1971). The photon indices have been obtained using \chandra\ {\sc pimms} v4.4 which also implements the correction required to account for the ACIS quantum-efficiency decay at low energies (Townsley \et 2000).}
\tablenotetext{c}{Count rate computed in the observed-frame \hbox{0.5--2~keV} band in units of $10^{-3}$~counts~s$^{-1}$.}
\tablenotetext{d}{Flux density at rest-frame 2~keV in units of $10^{-32}$~erg~cm$^{-2}$~s$^{-1}$~Hz$^{-1}$ assuming a power-law model with
$\Gamma=2.0$.}
\label{tab:chandra_counts}
\end{deluxetable*}

\subsection{Archival \xray\ Data}
\label{sec:archival_xray}

The data described in the previous two sections, for the \chandra\ and \swift\ sources, were complemented with archival observations from different \xray\ missions. Specifically, Table~\ref{tab:archival_log} provides an observation log of the archival, i.e., non-\swift/{\sl XRT}, \xray\ data for the \swift\ sources. For PG~1634$+$706, the {\sl Einstein} Imaging Proportional Counter (IPC) flux in the observed-frame \hbox{$0.28-3.29$~keV} band was obtained from Tananbaum \et (1986). Pointed {\sl ROSAT} Position Sensitive Proportional Counters observations for the three \swift\ sources were analyzed according to the prescription from HEASARC\footnote{http://heasarc.gsfc.nasa.gov/docs/rosat/ros\_xselect\_guide/\#tth\_sEc4.1} using
{\sc xselect} and {\sc ftools}. We screened the data for ``good'' events in the observed-frame $0.5-2.0$~keV band, and extracted the counts from source regions having a radius of $50''$ (this includes $\sim92$\% of the source flux at 1~keV in the observed frame); background counts were obtained from source-free regions having radii of $150''$. The net count rates are consistent with those obtained from the {\sl ROSAT} 2RXP catalog.\footnote{http://vizier.u-strasbg.fr/viz-bin/VizieR-3?-source=IX/30/2rxp} For the {\sl ASCA} observations of PG~1247$+$267 and PG~1634$+$706, we obtained fluxes in the observed-frame $2-10$~keV band from George \et (2000). The \xmm\ observations of all three \swift\ sources were processed using standard \xmm\ Science Analysis System\footnote{http://xmm.esac.esa.int/sas} v6.5.0 tasks. For each source, counts were extracted from the European Photon Imaging Camera pn detector using an aperture with a 30\arcsec\ radius centered on the source (this includes $\sim85$\% of the source flux at 1.5~keV in the observed frame); background counts were extracted from apertures of similar sizes in nearby source-free regions.
The \chandra\ observations of PG~1634$+$706 and HS~1700$+$6416 were analyzed in the same manner as described in
\S~\ref{sec:new_chandra} for the \chandra\ sources. Fluxes in the observed-frame $0.2-10$~keV band, corrected for Galactic absorption, for all these observations were obtained using WebPIMMS assuming a power-law photon index of $\Gamma=2.0$.

The archival \xray\ data for the \chandra\ sources from the first two epochs were obtained from the literature as indicated in
Table~\ref{tab:lc_chandra}. Unabsorbed fluxes in the observed-frame $0.5-2.0$~keV band were determined from the fluxes provided in the literature using WebPIMMS and assuming a power-law photon index of $\Gamma = 2.0$ (see also Shemmer \et 2005).

\begin{deluxetable*}{lclcccc}
\tablecolumns{7}
\tablecaption{Log of New \swift\ Observations of the \swift\ Sources}
\tablehead{
\colhead{} &
\colhead{} &
\colhead{} &
\colhead{Exp. Time} &
\colhead{Net {\sl XRT}} &
\colhead{{\sl UVOT}} &
\colhead{} \\
\colhead{Quasar} &
\colhead{Obs. ID} &
\colhead{Obs. Date} &
\colhead{(ks)} &
\colhead{Counts\tablenotemark{a}} &
\colhead{Filters} &
\colhead{Bin\tablenotemark{b}}
}
\startdata
PG~1247$+$267 & 00036675001 & 2007 Jun 17 & 2.89 & $53.5^{+8,4}_{-7.3}$ & $U, W2$ & 1 \\
                                & 00036674001 & 2007 Jun 18  & 3.73 & $65.8^{+9,2}_{-8.1}$ & $M2, W2$ & 1 \\
                                & 00090031001 & 2008 May 15 & 8.04 & $111.5^{+11,6}_{-10.5}$ & $V, U, B, W1, M2, W2$ & 2 \\
                                & 00090031002 & 2008 May 29 & 3.88 & $50.3^{+8,1}_{-7.1}$ & $V, U, B, W1, M2, W2$ & 3 \\
                                & 00090031003 & 2008 Nov 13 & 3.85 & $50.8^{+8,2}_{-7.1}$ & $V, U, B, W1, M2, W2$ & 4 \\
                                & 00036675002 & 2009 Feb 13 & 1.31 & $28.8^{+6,4}_{-5.3}$ & $U$ & 5 \\
                                & 00036676001 & 2009 Feb 14 & 5.71 & $80.2^{+10,0}_{-8.9}$ & $U$ & 5 \\
                                & 00091439001 & 2012 Dec 08 & 0.16 & \textless 4.8 & $W2$ & 6 \\
                                & 00091439002 & 2013 Jan 31 & 1.93 & $20.7^{+5,6}_{-4.5}$ & $W1$ & 6 \\
                                & 00091439003 & 2013 Feb 04 & 0.15 & \textless 4.8 & $W1$ & 6 \\
                                & 00091439004 & 2013 Mar 04 & 0.35 & $3.8^{+3,1}_{-1.9}$ & $W1$ & 6 \\
                                & 00091439005 & 2013 Mar 16 & 0.15 & \textless 4.8 & $W1$ & 6 \\
                                & 00091439006 & 2013 Mar 23 & 0.55 & $10.3^{+4,3}_{-3.2}$ & $M2$ & 6 \\
\hline
PG~1634$+$706 & 00036672001 & 2007 Jun 29 & 1.32 & $33.2^{+6,8}_{-5.7}$ & $U$ & 1 \\
                                & 00036673001 & 2007 Jun 29 & 1.48 & $44.8^{+7,8}_{-6.7}$ & $U$ & 1 \\
                                & 00036672002 & 2007 Jul 11 & 7.34 & $196.5^{+15,0}_{-14.0}$ & $U$ & 2 \\
                                & 00036671002 & 2008 Apr 22 & 2.09 & $94.3^{+10,8}_{-9.7}$ & $M2$ & 3 \\
                                & 00036673002 & 2008 Apr 24 & 2.56 & $96.2^{+10,8}_{-9.8}$ & $U$ & 3 \\
                                & 00036671003 & 2008 Apr 26 & 3.00 & $134.8^{+12,6}_{-11.6}$ & $M2$ & 4 \\
                                & 00090030001 & 2008 May 15 & 3.72 & $161.0^{+13,7}_{-12.7}$ & $V, U, B, W1, M2, W2$ & 5 \\
                                & 00090030002 & 2008 Jun 12 & 3.25 & $92.3^{+10,6}_{-9.6}$ & $V, U, B, W1, M2, W2$ & 6 \\
                                & 00090030003 & 2009 Jan 18 & 2.75 & $71.3^{+9,5}_{-8.4}$ & $V, U, B, W1, M2, W2$ & 7 \\
                                & 00091438001 & 2012 Jun 11 & 1.26 & $43.7^{+7,7}_{-6.6}$ & $W2$ & 8 \\
                                & 00091438002 & 2012 Jun 18 & 0.37 & $17.3^{+5,3}_{-4.1}$ & $U$ & 8 \\
                                & 00091438004 & 2012 Jun 22 & 0.42 & $18.3^{+5,4}_{-4.2}$ & $U,W2$ & 8 \\
                                & 00091438005 & 2012 Jun 25 & 0.17 & $7.7^{+3,9}_{-2.7}$ & $W1$ & 8 \\
                                & 00091438006 & 2012 Jun 26 & 0.36 & $23.3^{+5,9}_{-4.8}$ & $U$ & 8 \\
                                & 00091438007 & 2012 Jun 28 & 0.50 & $19.8^{+5,5}_{-4.4}$ & $M2$ & 8 \\
                                & 00091438008 & 2012 Jul 01 & 5.88 & $348.5^{+19,7}_{-18.7}$ & $M2,W2$ & 9 \\
\hline
HS~1700$+$6416 & 00036670001 & 2007 May 28 & 0.80 & \textless 3 & $U$ & 1 \\
                                  & 00036670002 & 2007 Jun 07 & 5.44 & $17.2^{+5.2}_{-4.1}$ & $M2$ & 1 \\
                                  & 00036669001 & 2007 Jun 27 & 4.92 & $6.0^{+3.6}_{-2.4}$ & $M2$ & 1 \\
                                  & 00036668001 & 2007 Jun 27 & 4.34 & $10.8^{+4.4}_{-3.2}$ & $M2$ & 2 \\
                                  & 00036669002 & 2008 Jan 17 & 2.10 & $11.7^{+4.5}_{-3.4}$ & $M2$ & 2 \\
                                  & 00036668002 & 2008 Apr 24 & 2.21 & $14.8^{+4.9}_{-3.8}$ & $U$ & 2 \\
                                  & 00090032001 & 2008 May 16 & 10.31 & $33.8^{+6.9}_{-5.8}$ & $V, U, B, W1, M2, W2$ & 3 \\
                                  & 00090032002 & 2008 May 30 & 4.64 & $20.7^{+5.6}_{-4.5}$ & $V, U, B, W1, M2, W2$ & 4 \\
                                  & 00090032003 & 2008 May 31 & 4.22 & $19.3^{+5.5}_{-4.4}$ & $V, U, B, W1, M2, W2$ & 4 \\
                                  & 00090032004 & 2009 Jan 21 & 1.99 & $6.2^{+3.6}_{-2.4}$ & $V, U, B, W1, M2, W2$ & 5 \\
                                  & 00090032005 & 2009 Jan 22 & 4.32 & $12.2^{+4.6}_{-3.4}$ & $V, U, B, W1, M2, W2$ & 5 \\
                                  & 00090032006 & 2009 Jan 29 & 0.50 & \textless 3 & $V, U, B, W1, M2, W2$ & 5 \\
                                  & 00090032007 & 2009 Jan 29 & 2.80 & $11.8^{+4.5}_{-3.4}$ & $V, U, B, W1, M2, W2$ & 5 \\
                                  & 00091440001 & 2012 Jul 03 & 0.75 & \textless 8 & $W1$ & 6 \\
                                  & 00091440002 & 2012 Jul 05 & 0.88 & $3.7^{+3.1}_{-1.8}$ & $W2$ & 6 \\
                                  & 00091440003 & 2012 Jul 06 & 4.15 & $19.3^{+5.5}_{-4.4}$ & $W1,M2$ & 6 \\
                                  & 00091440004 & 2012 Nov 23 & 0.82 & $6.3^{+3.7}_{-2.4}$ & $M2$ & 6 \\
                                  & 00091440006 & 2012 Nov 28 & 1.82 & $4.8^{+3.4}_{-2.1}$ & $W1$ & 6 \\
                                  & 00091440007 & 2012 Dec 09 & 1.47 & $6.2^{+3.6}_{-2.4}$ & $W1,M2$ & 7 \\
                                  & 00091440008 & 2012 Dec 30 & 1.39 & $4.5^{+3.3}_{-2.0}$ & $W1$ & 7 \\
                                  & 00091440009 & 2013 Jan 05 & 0.72 & \textless 6.4 & $W2$ & 7 \\
                                  & 00091440010 & 2013 Jan 08 & 0.87 & $4.2^{+3.2}_{-2.0}$ & $U$ & 7 \\
                                  & 00091440011 & 2013 Jan 09 & 0.92 & \textless 8 & $W2$ & 7 \\
                                  & 00091440012 & 2013 Jan 10 & 0.78 & \textless 8 & $M2$ & 7 \\
                                  & 00091440013 & 2013 Jan 12 & 4.06 & $20.8^{+5.6}_{-4.5}$ & $U$ & 7 \\
                                  & 00091440014 & 2013 Jan 13 & 3.55 & $17.5^{+5.3}_{-4.1}$ & $M2,W2$ & 8 \\
                                  & 00091440015 & 2013 Jan 15 & 1.50 & $4.7^{+3.3}_{-2.1}$ & $W1$ & 8 \\
                                  & 00091440016 & 2013 Jan 16 & 0.80 & $5.8^{+3.6}_{-2.3}$ & $U$ & 8
\enddata
\tablenotetext{a}{Errors on the \xray\ counts were computed according to Tables~1 and 2 of Gehrels (1986) and correspond to the 1$\sigma$ level; these were calculated using Poisson statistics and include careful background subtraction. The upper limits are at the 95\% confidence level and were computed according to Kraft \et (1991). Upper limits of 3.0, 4.8, 6.4, and 8.0 indicate that 0, 1, 2, and 3 \xray\ counts, respectively, have been found within an extraction region of radius 20 pixels centered on the optical position of the quasar (assuming the background within this source-extraction region to be negligible).}
\tablenotetext{b}{\swift\ observations having the same bin number were stacked in order to compute the \xray\ variability probability and amplitude.}
\label{tab:swift_log}
\end{deluxetable*}

\section{RESULTS}
\label{sec:results}

\subsection{Variability Amplitudes}
\label{sec:results_amplitudes}

\begin{figure}
\plotone{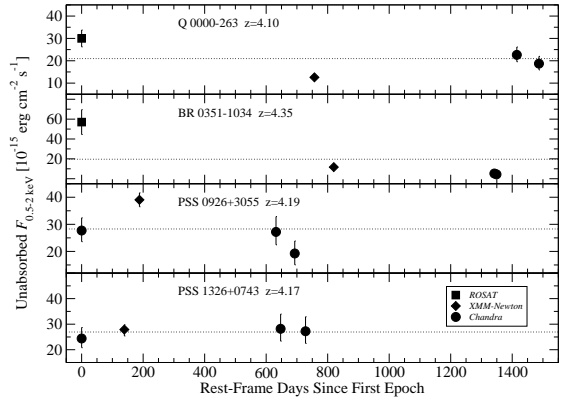}
\caption{\xray\ light curves of the \chandra\ sources. Galactic-absorption corrected flux in the observed-frame $0.5-2$~keV band is plotted as a function of rest-frame time (in days) relative to the first \xray\ epoch for each source. Squares, diamonds, and circles mark {\sl ROSAT}, \xmm, and \chandra\ observations, respectively. The dotted line in each panel indicates the mean flux.}
\label{fig:LC_Chandra}
\end{figure}

\begin{figure}
\plotone{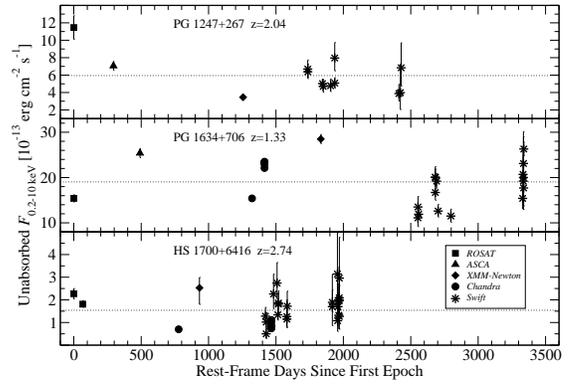}
\caption{\xray\ light curves of the \swift\ sources. Galactic-absorption corrected flux in the observed-frame $0.2-10$~keV band is plotted as a function of rest-frame time (in days) relative to the first \xray\ epoch for each source. Squares, triangles, diamonds, circles, and stars indicate {\sl ROSAT},
{\sl ASCA}, \xmm, \chandra, and \swift\ observations, respectively. The much earlier {\sl Einstein} observation of PG~1634$+$706 with an unabsorbed flux of \hbox{$F_{\rm 0.2 - 10~keV} = (16.0\pm3.2) \times 10^{-13}$~erg~cm$^{-2}$~s$^{-1}$} is not shown in the middle panel, for clarity. The dotted line in each panel indicates the mean flux.}
\label{fig:LC_Swift}
\end{figure}

Tables~\ref{tab:lc_chandra}~and~\ref{tab:lc_swift} provide the \xray\ temporal data of the \chandra\ and \swift\ sources, respectively. The respective light curves are presented in Figures~\ref{fig:LC_Chandra}~and~\ref{fig:LC_Swift}. The first step in the time-series analysis is to determine whether a source is variable. This is performed by applying a $\chi^2$ test to the light curve of each source, where the null hypothesis is that the flux in each epoch is consistent with the mean flux of the entire light curve, within the errors. This can be written as

\begin{equation}
\chi^2={1\over N_{\rm obs}-1}\sum_{i=1}^{N_{\rm obs}}\frac{(f_i-\langle f \rangle)^2}{\sigma_i^2}
\end{equation}

\noindent where $f_i$ and $\sigma_i$ are the flux and its error in the {\it i}th observation, respectively, $N_{\rm obs}$ is the number of observations, and $\langle f \rangle$ is the mean flux of the light curve (e.g., P04). Table~\ref{tab:variability} gives the $\chi^2$ values as well as the corresponding degrees of freedom (dof; i.e., $N_{\rm obs} - 1$) and the $\chi^2$ distribution probability by which the null hypothesis can be rejected ($1-p$).

Due to \swift's planning and scheduling constraints, our requested exposures of the \swift\ sources were, in most cases, divided into shorter
exposure segments with uneven integration times as can be seen in Table~\ref{tab:swift_log}. For the purpose of determining whether a \swift\
source is variable and to compute its variability amplitude, we stacked the counts in the \swift\ exposures into separate bins as indicated in
Table~\ref{tab:swift_log}. By attempting to match our original requested exposures, we binned the \swift\ data by stacking exposures, chronologically, up to a minimum of 4~ks, 3~ks, and 8~ks, rounded to the nearest  ks, for PG~1247$+$267, PG~1634$+$706, and HS~1700$+$6416, respectively. However, we stacked exposures from Cycle~8 separately from the other Cycles due to the large time gap up to Cycle~8. The original requested exposures were designed to provide uncertainties of $\approx10-20$\% on the flux measurements of the \swift\ sources. The $\chi^2$ values and probabilities for the \swift\ sources shown in Table~\ref{tab:variability} were computed based on their binned \swift\ data.

Since the light curves of our sources are composed of data from different \xray\ observatories, flux uncertainties stemming from instrument cross calibrations may affect the measured variability properties. One particular concern is that the two \chandra\ sources that have three \chandra\ observations and one observation with \xmm\ appear to be either non-variable (PSS~1326$+$0743) or much less variable (PSS~0926$+$3055) than the other two \chandra\ sources that have two observations with \chandra, one with {\sl ROSAT}, and one with \xmm. This trend may be a consequence of real larger variations due to the longer temporal baselines introduced by the {\sl ROSAT} observations (as discussed further below),
or due to the fact that data from multiple observatories introduce additional cross-calibration uncertainties. Sources may thus appear more variable if these uncertainties are not properly taken into account. It underscores the importance of conducting any monitoring project in a uniform way with the same instrument in the same band in order to obtain the most reliable results.

The relative uncertainties in the cross calibrations among the \xray\ observatories used in this work (except for {\sl Einstein/IPC}) are typically of the order of $\ltsim10$\% (e.g., Kirsch \et 2005). After adding, in quadrature, an uncertainty of 10\% to the error bar on each flux value to account for these uncertainties (following Saez \et 2012), the significance of the $\chi^2$ test drops below 90\% for each of the \swift\ sources and for PSS~0926$+$3055. For the \swift\ sources, this effect is mainly due to the fact that the S/N of their \swift\ data are lower, on average, than those of the other \xray\ data and their sampling pattern is more complex, relying heavily on the shortest sampling timescales where uncertainties on the variability power are larger; as discussed below, the variability amplitude increases with sampling timescale.

To overcome the sampling bias, and to allow a better comparison with the temporal patterns of the \chandra\ sources, we computed the $\chi^2$ value for each \swift\ source, including the additional cross-calibration uncertainties of 10\%, and using a light curve composed of one data point per \xray\ observatory, where the fluxes from each observatory were averaged (resulting in a substantial increase in the S/N of the \swift\ data).
As a result, PG~1247$+$267 and HS~1700$+$6416 are variable with $>90$\% confidence, whereas PG~1634$+$706 shows variability
with $>90$\% confidence only when the first (i.e., {\sl Einstein}) and last (i.e., \swift) data points are excluded. The removal of these two points was
done in order to match, as closely as possible, the temporal sampling of the other \swift\ and \chandra\ sources, and due to the fact that the cross-calibration uncertainty of the {\sl Einstein/IPC} with respect to the other instruments we use has not yet been determined (Kirsch \et 2005).
In particular,  the rest-frame time separations between the {\sl Einstein} epoch and the {\sl ROSAT} epoch (i.e., second epoch) as well as between
the mean epoch of the \swift\ observations and the \xmm\ epoch (i.e., fifth epoch) are a factor of $\sim2-3$ larger than the average and median rest-frame time separations between adjacent epochs in all our sources.

The light curve of PSS~0926$+$3055 is composed of three \chandra\ epochs and one \xmm\ epoch, and the relative calibration uncertainties between \chandra\ and \xmm\ are considerably lower than 10\% (e.g., Snowden \et 2002). We therefore maintain that small cross-calibration uncertainties do not warrant an artificial increase in the flux uncertainties for this source. In fact, when the added cross-calibration uncertainty is 3\%, the $\chi^2$ significance $>90$\%. To summarize, we consider all our sources, except for PSS~1326$+$0743, to be \xray\ variable at $>90$\% confidence.

The \xray\ variability amplitude is computed as the source excess variance defined by Turner \et (1999; see also Nandra \et 1997) as

\begin{equation}
\sigma^2_{\rm rms} = {1 \over N_{\rm obs} \langle f \rangle ^2} \sum_{i=1}^{N_{\rm obs}} \left [ \left (f_i - \langle f \rangle \right )^2 - \sigma_i^2 \right ],
\end{equation}

\noindent who also define the error on $\sigma^2_{\rm rms}$ as $s_D / (\langle f \rangle ^2 \sqrt{N_{\rm obs}})$, where $s_D$ is given by

\begin{equation}
s^2_D = {1 \over N_{\rm obs}-1} \sum_{i=1}^{N_{\rm obs}} \left \{ \left [ \left (f_i - \langle f \rangle \right )^2 - \sigma_i^2 \right ] - \sigma^2_{\rm rms}\langle f \rangle ^2 \right \}^2.
\end{equation}

\noindent The $\sigma^2_{\rm rms}$ values and their errors for our sources are listed in Table~\ref{tab:variability}; for the \swift\ sources, these were computed using the binned \swift\ observations. The above expression for the error on $\sigma^2_{\rm rms}$, however, represents only a `formal' error which does not take into account the scatter intrinsic to any red-noise random process, as discussed in Vaughan \et (2003) and Allevato \et (2013). The quantity $\sigma^2_{\rm rms}$ is also prone to systematic effects and biases stemming from the duration of the monitoring and from the power-law slope of the AGN power spectral density (PSD) function. The {\em intrinsic} $\sigma^2_{\rm rms}$ tends to increase with the monitoring duration (e.g., Vagnetti \et 2011) due to the red-noise nature of the PSD function, while the {\em observed} $\sigma^2_{\rm rms}$ increases with steeper PSD power-law slopes with respect to the intrinsic value due to power leakage from lower frequencies (Allevato \et 2013).

\begin{deluxetable*}{llllcl}
\tablecolumns{6}
\tablecaption{Log of Archival Observations of the \swift\ Sources}
\tablehead{
\colhead{} &
\colhead{} &
\colhead{} &
\colhead{} &
\colhead{Exp. Time} &
\colhead{} \\
\colhead{Quasar} &
\colhead{Observatory} &
\colhead{Obs. ID} &
\colhead{Obs. Date} &
\colhead{(ks)} &
\colhead{References}
}
\startdata
PG~1247$+$267 & {\sl ROSAT} & 701173           & 1993 Jan 04 & 2.10 & 1 \\
                                & {\sl ASCA}    & 73048000      & 1995 Jun 17 & 35.93 & 2 \\
                                & \xmm\            & 0143150201 & 2003 Jun 18 & 33.92 & 1, 3 \\
\hline
PG~1634$+$706 & {\sl Einstein} & I05351 & 1981 Feb 6 & 1.83 & 4 \\
                                & {\sl ROSAT} & 700246  & 1991 Mar 15 &  9.01 & 1 \\
                                & {\sl ASCA}    & 71036000 & 1994 May 02 & 40.54 & 2 \\
                                & \chandra\      & \dataset [ADS/Sa.CXO#obs/1269] {1269} & 1999 Aug 21 & 10.83 & 1 \\
                                & \chandra\      & \dataset [ADS/Sa.CXO#obs/47] {47} & 2000 Mar 23 & 5.39 & 1 \\
                                & \chandra\      & \dataset [ADS/Sa.CXO#obs/62] {62} & 2000 Mar 23 & 4.85 & 1 \\
                                & \chandra\      & \dataset [ADS/Sa.CXO#obs/69] {69} & 2000 Mar 24 & 4.86 & 1 \\
                                & \chandra\      & \dataset [ADS/Sa.CXO#obs/70] {70} & 2000 Mar 24 & 4.86 & 1 \\
                                & \chandra\      & \dataset [ADS/Sa.CXO#obs/71] {71} & 2000 Mar 24 & 4.41 & 1 \\
                                & \xmm\            & 0143150101 & 2002 Nov 22 & 19.71 & 1, 3, 5\\
\hline
HS~1700$+$6416  & {\sl ROSAT} & 701121          & 1992 Nov 13 & 15.50 & 1,6 \\
                                   & {\sl ROSAT} & 701457           & 1993 Jul 21  & 26.32 & 1,6 \\
                                   & \chandra\      & \dataset [ADS/Sa.CXO#obs/547] {547}                 & 2000 Oct 31 & 49.53 & 1, 7, 8, 9 \\
                                   & \xmm\            & 0107860301 & 2002 May 31 & 27.10 & 1, 7, 8, 9 \\
                                   & \chandra\      & \dataset [ADS/Sa.CXO#obs/8032] {8032}               & 2007 Nov 12 & 31.00 & 1, 9 \\ 
                                   & \chandra\      & \dataset [ADS/Sa.CXO#obs/9757] {9757}               & 2007 Nov 13 & 20.78 & 1, 9 \\
                                   & \chandra\      & \dataset [ADS/Sa.CXO#obs/9756] {9756}               & 2007 Nov 14 & 32.26 & 1, 9 \\
                                   & \chandra\      & \dataset [ADS/Sa.CXO#obs/9758] {9758}               & 2007 Nov 16 & 23.37 & 1, 9 \\
                                   & \chandra\      & \dataset [ADS/Sa.CXO#obs/9759] {9759}               & 2007 Nov 17 & 31.19 & 1, 9 \\
                                   & \chandra\      & \dataset [ADS/Sa.CXO#obs/9760] {9760}               & 2007 Nov 19 & 16.95 & 1, 9 \\
                                   & \chandra\      & \dataset [ADS/Sa.CXO#obs/8033] {8033}               & 2007 Nov 20 & 29.71 & 1, 9 \\
                                   & \chandra\      & \dataset [ADS/Sa.CXO#obs/9767] {9767}               & 2007 Nov 21 & 9.04 & 1, 9
\enddata
\tablerefs{(1) this work; (2) George \et (2000); (3) Page \et (2004); (4) Tananbaum \et (1986); (5) Piconcelli \et (2005); (6) Reimers \et (2005); (7) Just \et (2007); (8) Misawa \et (2008); (9) Lanzuisi \et (2012).}
\label{tab:archival_log}
\end{deluxetable*}

\begin{deluxetable}{lcccl}
\tablecolumns{5}
\tablecaption{X-ray Light Curve Data for the \chandra\ Sources}
\tablehead{
\colhead{} &
\colhead{} &
\colhead{} &
\colhead{} &
\colhead{} \\
\colhead{Quasar} &
\colhead{JD} &
\colhead{$f_{\rm x}$\tablenotemark{a}} &
\colhead{Observatory} &
\colhead{Reference}
}
\startdata
Q~0000$-$263 & 2448588.5 & 30.0$\pm$3.8 & {\sl ROSAT} & 1, 2, 3 \\
& 2452450.5 & 12.6$\pm$0.7 & \xmm\ & 3, 4, 5 \\
& 2455802.5 & 22.6$^{+3.5}_{-3.1}$ & \chandra\ & 6 \\
& 2456173.5 & 18.7$^{+3.2}_{-2.8}$ & \chandra\ & 6 \\
BR~0351$-$1034 & 2448647.5 & 57.0$\pm$12.5 & {\sl ROSAT} & 2, 3 \\
& 2453035.5 & 11.7$\pm$2.2 & \xmm\ & 3, 5, 7 \\
& 2455827.5 & 5.3$^{+2.0}_{-1.5}$ & \chandra\ & 6 \\
& 2455862.5 & 4.4$^{+1.9}_{-1.4}$ & \chandra\ & 6 \\
PSS~0926$+$3055 & 2452344.5 & 27.7$^{+4.7}_{-4.1}$ & \chandra\ & 3, 8 \\
& 2453322.5 & 39.0$^{+2.6}_{-2.5}$ & \xmm\ & 3 \\
& 2455623.5 & 27.2$^{+5.7}_{-4.7}$ & \chandra\ & 6 \\
& 2455939.5 & 19.2$^{+4.6}_{-4.2}$ & \chandra\ & 6 \\
PSS~1326$+$0743 & 2452284.5 & 24.4$^{+4.3}_{-3.6}$ & \chandra\ & 3, 8 \\
& 2453001.5 & 27.9$^{+1.9}_{-2.5}$ & \xmm\ & 3 \\
& 2455627.5 & 28.2$^{+5.8}_{-4.8}$ & \chandra\ & 6 \\
& 2456047.5 & 27.2$^{+5.7}_{-4.8}$ & \chandra\ & 6
\enddata
\tablenotetext{a}{Galactic absorption-corrected flux in the observed-frame \hbox{0.5--2~keV} band in units of $10^{-15}$~erg~cm$^{-2}$~s$^{-1}$.}
\tablerefs{(1) Bechtold \et (1994); (2) Kaspi \et (2000b); (3) Shemmer \et (2005); (4) Ferrero \& Brinkmann (2003); (5) Grupe \et (2006); (6) this work; (7) Grupe \et (2004); (8) Vignali \et (2003).}
\label{tab:lc_chandra}
\end{deluxetable}

\tabletypesize{\scriptsize}

\begin{deluxetable*}{llccc}
\tablecolumns{5}
\tablecaption{X-ray Light Curve Data for the \swift\ Sources}
\tablehead{
\colhead{} &
\colhead{} &
\colhead{} &
\colhead{Observatory} &
\colhead{} \\
\colhead{Quasar} &
\colhead{JD} &
\colhead{$f_{\rm x}$\tablenotemark{a}} &
\colhead{Code} &
\colhead{\aox}
}
\startdata
PG~1247$+$267 & 2448991.5 & $11.5\pm1.4$ & 1 & \nodata \\
& 2449886.0 & $7.1\pm0.5$ & 2 & \nodata \\
& 2452808.9 & $3.5\pm0.2$ & 3 & \nodata \\
& 2454268.5 & $6.7^{+1.0}_{-0.9}$ & 4 & $-1.65$ \\
& 2454269.8 & $6.4^{+0.9}_{-0.8}$ & 4 & $-1.64$ \\
& 2454601.5 & $5.0\pm0.5$ & 4 & $-1.70$ \\
& 2454615.9 & $4.7^{+0.8}_{-0.7}$ & 4 & $-1.72$ \\
& 2454783.7 & $4.8^{+0.8}_{-0.7}$ & 4 & $-1.71$ \\
& 2454876.5 & $8.0^{+1.8}_{-1.5}$ & 4 & $-1.64$ \\
& 2454877.1 & $5.1\pm0.6$ & 4 & $-1.72$ \\
& 2456323.5 & $3.9^{+1.1}_{-0.8}$ & 4 & $-1.75$ \\
& 2456356.5 & $4.0^{+3.3}_{-1.9}$ & 4 & $-1.75$ \\
& 2456375.1 & $6.8^{+2.9}_{-2.1}$ & 4 & $-1.66$ \\
\hline
PG~1634$+$706 & 2444641.5 & 16.0$\pm$3.2 & 5 & \nodata \\
& 2448330.9 & 15.4$\pm$0.1 & 1 & \nodata \\
& 2449475.3 & 25.4$\pm$1.0 & 2 & \nodata \\
& 2451412.5 & $15.4\pm0.3$ & 6 & \nodata \\
& 2451627.4 & $23.3\pm0.6$ & 6 & \nodata \\
& 2451627.5 & $23.4\pm0.6$ & 6 & \nodata \\
& 2451627.5 & $22.9\pm0.6$ & 6 & \nodata \\
& 2451627.6 & $22.1\pm0.6$ & 6 & \nodata \\
& 2451627.7 & $23.5^{+0.7}_{-0.6}$ & 6 & \nodata \\
& 2452601.4 & $28.5\pm1.0$ & 3 & \nodata \\
& 2454280.5 & $11.1^{+2.3}_{-1.9}$ & 4 & $-1.78$ \\
& 2454280.9 & $13.5^{+2.3}_{-2.0}$ & 4 & $-1.75$ \\
& 2454292.5 & $11.9^{+0.9}_{-0.8}$ & 4 & $-1.77$ \\
& 2454579.3 & $20.1^{+2.3}_{-2.1}$ & 4 & $-1.68$ \\
& 2454581.0 & $16.7^{+1.9}_{-1.7}$ & 4 & $-1.71$ \\
& 2454582.6 & $20.0^{+1.9}_{-1.7}$ & 4 & $-1.68$ \\
& 2454601.9 & $19.2^{+1.6}_{-1.5}$ & 4 & $-1.69$ \\
& 2454629.5 & $12.6^{+1.5}_{-1.3}$ & 4 & $-1.76$ \\
& 2454850.1 & $11.5^{+1.5}_{-1.4}$ & 4 & $-1.77$ \\
& 2456090.0 & $15.4^{+2.7}_{-2.3}$ & 4 & $-1.68$ \\
& 2456096.8 & $20.7^{+6.3}_{-4.9}$ & 4 & $-1.65$ \\
& 2456100.8 & $19.2^{+5.6}_{-4.4}$ & 4 & $-1.64$ \\
& 2456103.9 & $20.0^{+10.2}_{-7.0}$ & 4 & $-1.65$ \\
& 2456104.6 & $23.1^{+5.9}_{-4.8}$ & 4 & $-1.64$ \\
& 2456107.4 & $17.7^{+4.9}_{-3.9}$ & 4 & $-1.66$ \\
& 2456110.2 & $26.3^{+1.5}_{-1.4}$ & 4 & $-1.60$ \\
\hline
HS~1700$+$6416 & 2448939.5 & $2.3\pm0.2$ & 1 & \nodata \\
& 2449189.5 & $1.8\pm0.2$ & 1 & \nodata \\
& 2451849.5 & 0.70$\pm$0.04 & 6 & \nodata \\
& 2452426.3 & $2.5^{+0.5}_{-0.7}$ & 3 & \nodata \\
& 2454258.7 & $1.3^{+0.4}_{-0.3}$ & 4 & $-1.89$ \\
& 2454278.6 & $0.5^{+0.3}_{-0.2}$ & 4 & $-2.04$ \\
& 2454279.1 & $1.0^{+0.4}_{-0.3}$ & 4 & $-1.92$ \\
& 2454416.6 & $0.8\pm0.1$ & 6 & \nodata \\
& 2454418.2 & $0.7\pm0.1$ & 6 & \nodata \\
& 2454419.3 & $0.8\pm0.1$ & 6 & \nodata \\
& 2454420.9 & $1.0\pm0.1$ & 6 & \nodata \\
& 2454421.9 & $0.8\pm0.1$ & 6 & \nodata \\
& 2454423.6 & $1.1\pm0.1$ & 6 & \nodata \\
& 2454424.6 & $0.9\pm0.1$ & 6 & \nodata \\
& 2454425.9 & $0.8\pm0.1$ & 6 & \nodata \\
& 2454482.7 & $2.3^{+0.9}_{-0.7}$ & 4 & $-1.79$ \\
& 2454580.5 & $2.7^{+0.9}_{-0.7}$ & 4 & $-1.75$ \\
& 2454602.5 & $1.3^{+0.3}_{-0.2}$ & 4 & $-1.88$ \\
& 2454617.0 & $1.8^{+0.5}_{-0.4}$ & 4 & $-1.82$ \\
& 2454618.0 & $1.9^{+0.5}_{-0.4}$ & 4 & $-1.82$ \\
& 2454852.5 & $1.3^{+0.7}_{-0.5}$ & 4 & $-1.88$ \\
& 2454853.5 & $1.2^{+0.4}_{-0.3}$ & 4 & $-1.89$ \\
& 2454861.5 & $1.7^{+0.7}_{-0.5}$ & 4 & $-1.81$ \\
& 2456113.5 & $1.7^{+1.4}_{-0.8}$ & 4 & $-1.84$ \\
& 2456115.1 & $1.9^{+0.5}_{-0.4}$ & 4 & $-1.82$ \\
& 2456254.6 & $3.1^{+1.8}_{-1.2}$ & 4 & $-1.72$ \\
& 2456259.6 & $1.1^{+0.8}_{-0.5}$ & 4 & $-1.91$ \\
& 2456271.3 & $1.7^{+1.0}_{-0.7}$ & 4 & $-1.83$ \\
& 2456291.6 & $1.3^{+1.0}_{-0.6}$ & 4 & $-1.87$ \\
& 2456300.6 & $2.0^{+1.5}_{-0.9}$ & 4 & $-1.81$ \\
& 2456304.7 & $2.1^{+0.6}_{-0.5}$ & 4 & $-1.80$ \\
& 2456305.8 & $2.0^{+0.6}_{-0.5}$ & 4 & $-1.83$ \\
& 2456307.8 & $1.3^{+0.9}_{-0.6}$ & 4 & $-1.89$ \\
& 2456308.5 & $3.0^{+1.8}_{-1.2}$ & 4 & $-1.74$
\enddata
\tablecomments{Observatory Codes: (1) {\sl ROSAT}; (2) {\sl ASCA}; (3) \xmm; (4) \swift; (5)  {\sl Einstein}; (6) \chandra.}
\tablenotetext{a}{Galactic absorption-corrected flux in the observed-frame \hbox{0.2--10~keV} band in units of $10^{-13}$~erg~cm$^{-2}$~s$^{-1}$.}
\label{tab:lc_swift}
\end{deluxetable*}

\subsubsection{Comparison with the CDF-S 2~Ms Survey}
\label{sec:CDFS_2Ms}

We compare the variability amplitudes of our sources with those of \xray-selected AGN from the 2~Ms exposure of the \hbox{CDF-S} survey (Luo \et 2008). This sample forms an extension of the P04 variability study based on the 1~Ms exposure of the CDF-S and is a subset of the investigation based on its 4~Ms exposure (M.~Paolillo et al., in prep.). The comparison with the data from the 2~Ms exposure, instead of the P04 data or with the full 4~Ms exposure, is motivated by the fact that the timescales probed in this survey, spanning $\sim2-7$~yr in the rest frame, closely match the timescales probed for our \chandra\ and \swift\ sources. A distribution of the rest-frame timescales probed for sources in the Luo \et (2008) CDF-S survey is shown in Fig.~\ref{fig:timescales_2Ms}. The analysis of the \hbox{CDF-S} data was performed following the steps outlined in P04. The current study includes 81 point sources with \hbox{$L_{0.5-8~\rm keV}>10^{42}$}~erg~s$^{-1}$ and with $>300$ counts in order to meet the requirement of having $\sim10$ counts, on average, per \chandra\ observation. We further removed 20 of these sources that have radio detections (Xue \et 2011) in order to minimize potential jet-related variability. We do not imply, however, that all the remaining sources are formally radio quiet or that no RQQs were culled in this process; a rigorous analysis of the radio and optical properties of the CDF-S sources is beyond the scope of this work. Of the remaining 61 sources, 13 are deemed non variable, as their $\chi^2$ values indicate variability with $<90$\% confidence; yet, these non-variable sources are taken into account in the analysis below in order to prevent biasing of the variability-amplitude distribution (see, e.g., P04).

In Fig.~\ref{fig:Paolillo04_fig12} we plot the variability amplitudes of our sources as a function of \xray\ luminosity and show for reference the 61
\hbox{CDF-S} sources (cf. Fig.~12 of P04). To match the \xray\ luminosities of the \hbox{CDF-S} sources, the \xray\ luminosities of our sources were extrapolated to the observed-frame $0.5-8$~keV band by assuming a photon index of $\Gamma = 2.0$ for each source in this band. Our sources extend the \hbox{CDF-S} parameter space by $\Delta z \simeq 1$ and by an order of magnitude in \xray\ luminosity. Fig.~\ref{fig:Paolillo04_fig12} also shows simulated variability amplitudes as a function of luminosity which are described  in \S~\ref{sec:discussion}. By inspection of
Fig.~\ref{fig:Paolillo04_fig12}, one can see that extremely luminous AGN at the highest accessible redshifts continue to display \xray\ variability amplitudes that are as high or even higher than those of many of their lower-luminosity counterparts. We do not see a trend of decreasing variability amplitude with increasing luminosity as might have been expected based on studies of nearby AGNs (see \S~\ref{sec:introduction}). Our findings therefore bolster the results of P04 and support previous tentative reports of increased \xray\ variability in luminous AGN. However, as we discuss below, we do not find clear evidence that the \xray\ variability amplitude depends on redshift.

\subsection{Variability Timescales}
\label{sec:sf}

Although the \chandra\ sources lack sufficient observations for a quantitative temporal analysis, we detect significant \xray\ variability over characteristic rest-frame timescales of $\approx10^2 - 10^3$~d in the variable sources. No \xray\ variability is detected in these sources over rest-frame timescales of $\ltsim0.1$~d (Shemmer \et 2005). Additional \chandra\ observations are required for a quantitative analysis, especially on intermediate rest-frame timescales ($\approx 1 - 10$~d). The \swift\ sources, on the other hand, have sufficient observations for a more detailed investigation of \xray\ variability timescales using variability structure functions (SFs). Even though the SF is not as sensitive to the variability power at each timescale as the PSD function (e.g., Emmanoulopoulos \et 2010), it is the best analysis tool available for sparsely sampled light curves with few epochs (e.g., Vagnetti \et 2011).

We computed the SF, i.e., $\Delta m$, for each \swift\ source following the definition from Fiore \et (1998):

\begin{equation}
\Delta m_{ji} = |  {2.5  \log \left [ f(t_j)/f(t_i) \right ] }  | 
\end{equation}

\noindent where $f(t_j)$ and $f(t_i)$ are the fluxes of the source at epochs $t_j$ and  $t_i$, respectively, such that $t_j > t_i$, and every $t_i$ is measured in rest-frame days since the first epoch (i.e., $t_1 = 0$). We binned the SF of each source into 13 rest-frame time bins forming a geometric sequence of the form \hbox{$(\Delta t_{\rm rest})_n = 2^{n-1}$~d} where $n\in1,\cdots,13$, and computed the mean SF,
$\langle \Delta m \rangle$, for each of these time bins.

Figure~\ref{fig:SF_Swift} shows the mean SFs of our \swift\ sources. The SFs show a general trend of increasing variability on longer rest-frame timescales ($\gtsim 100$~d), although variability at some level is detected in all the \swift\ sources at much shorter rest-frame timescales, down to
$\sim1$~d, in particular, in HS~1700$+$6416 (see Table~\ref{tab:lc_swift}). However, at the shortest timescale, the SF may be dominated by the measurement noise (e.g., Hughes \et 1992). We also computed an ensemble SF including all the \swift\ sources, by averaging the $\Delta m$ values of all the sources in each rest-frame time bin. This ensemble SF is plotted in Figure~\ref{fig:SF_compare} against the ensemble SFs of steep- and flat-\xray-spectrum quasars from Fiore \et (1998) that have luminosities that are almost two orders of magnitude lower than those of the \swift\ sources. The ensemble SF of the \swift\ sources appears to be remarkably similar to the ensemble SF of the steep-\xray\ spectrum quasars, which also have higher accretion rates (Fiore \et 1998), at least on the shorter timescales ($\ltsim10$\,d). Figure~\ref{fig:SF_compare} also shows that additional \xray\ monitoring is required to characterize better the ensemble SF on rest-frame timescales of $\sim10-100$~d and to compare with the behaviors of the \xray\ variability SFs from Vagnetti \et (2011). Finally, we searched for rapid \xray\ variability, on timescales of a few minutes in the rest frame, in the \swift\ sources by binning in time each \swift\ observation with more than 100 source counts. We did not find evidence for such rapid variability in the only two sources that have more than this number of counts in a single observation, PG~1247$+$267 and PG~1634$+$706 (see also Lanzuisi \et 2012 for the lack of rapid variability, over timescales of minutes, in the \chandra\ observations of HS~1700$+$6416).

\subsection{SED Variability}
\label{sec:aox}

AGNs often display complex connections between \xray\ variability and variations in the fluxes of other spectral bands due to the nature of the corresponding emission regions and the different PSD functions involved (see, e.g., Uttley \& McHardy 2004 and references therein). We search for connections between the \xray\ and optical-UV variations of our sources by obtaining optical-UV observations that are simultaneous (using
\swift/{\sl UVOT}) or nearly simultaneous (using ground-based observations) with the \xray\ observations. Specifically, the optical-UV data allow searching for variations in \aox\ (defined in Table~\ref{tab:chandra_props}), considered to be a measure of the accretion-disk corona reprocessing of optical-UV radiation emitted from the accretion disk. Significant \aox\ variations can probe the interplay between the accretion disk and the corona and provide insight into the accretion processes in the high-redshift sources under study.

\subsubsection{\swift/{\sl UVOT}  Observations}
\label{sec:uvot}

The {\sl UVOT} data for the \swift\ sources were extracted from the Italian Space Agency Science Data
Center\footnote{http://www.asdc.asi.it/mmia/index.php?mission=swiftmastr} archive. Flux densities in each {\sl UVOT} band were obtained by employing standard {\sc heasoft}\footnote{http://heasarc.gsfc.nasa.gov/docs/software/lheasoft/} tasks on the {\sl UVOT} sky images. The {\sl UVOT} filters available for each \swift\ observation are listed in Table~\ref{tab:swift_log}. For each \swift\ source, the flux density at a rest-frame wavelength of 2500~\AA, $f_{\nu}$(2500~\AA), was obtained by extrapolating the flux density available for the band with an effective wavelength which is closest to 2500~\AA\ using a power-law continuum of the form $f_{\nu} \propto \nu^{-0.5}$ (Vanden Berk \et 2001). These bands were either $V$, $B$, or $U$; hence the rest-frame effective wavelength for each observation, from which the $f_{\nu}$(2500~\AA) value was extrapolated, was typically
$\gtsim1200$~\AA. For observations lacking $V$, $B$, and $U$ filters, the flux densities from one of the other three available UV filters were converted to $V$-band fluxes using the mean flux-density ratio between that filter and the $V$ filter, based on available observations of the same source that included both the UV filter and the $V$ filter. The resulting \aox\ values are given in Table~\ref{tab:lc_swift}.

The largest differences in \aox\ between two \swift\ observations are \hbox{\daox$=0.11, 0.18, {\rm and}~0.32$} for PG~1247$+$267,
PG~1634$+$706, and HS~1700$+$6416, respectively; however, only for PG~1634$+$706 is this difference statistically significant (at $>3\,\sigma$). For all three \swift\ sources, the differences in \aox\ are dominated by the \xray\ variations (as shown in Table~\ref{tab:lc_swift}). At least for
PG~1634$+$706, these variations in \aox\ are larger than the $1~\sigma$ uncertainty on \aox\ values measured from the the correlation between \aox\ and the optical-UV luminosity for such high-luminosity AGNs (Steffen \et 2006). This result indicates that \xray\ variability may provide a significant contribution to the dispersion in the luminosity-corrected \aox\  values of luminous AGNs. Our results are consistent with the findings of Vagnetti \et (2013) for a low-redshift AGN sample observed with \swift\ (see also Gibson \et 2008; Vagnetti \et 2010; Gibson \& Brandt 2012). Although a detailed optical-UV-\xray\ cross-correlation analysis is beyond the scope of this work, the dominance of the \xray\ variations in the variations of \aox\ is also manifested by the considerably lower optical amplitudes of all three \swift\ sources during periods that overlap with our monitoring. The amplitudes of the optical continuum variations of these sources are normally within $\sim10$\% (see Kaspi \et 2007; Trevese \et 2007), while the typical \xray\ amplitudes are much larger, up to factors of a few.

\tabletypesize{\footnotesize}

\begin{deluxetable}{llccc}
\tablecolumns{5}
\tablecaption{X-ray Variability Indicators}
\tablehead{
\colhead{Sample} &
\colhead{Quasar} &
\colhead{$\chi^2$(dof)} &
\colhead{$1-p$\tablenotemark{a}} &
\colhead{$\sigma^2_{\rm rms}$}
}
\startdata
\chandra\ & Q~0000$-$263         & 46.4(3)    & $4.8 \times 10^{-10}$ & 0.07$\pm$0.05 \\
                  & BR~0351$-$1034    & 58.6(3)    & $1.2 \times 10^{-12}$ & 1.13$\pm$0.71 \\
                  & PSS~0926$+$3055 & 7.3(3)      & $6.2 \times 10^{-2}$   & 0.04$\pm$0.04 \\
                  & PSS~1326$+$0743 & 0.2(3)      & $9.7 \times 10^{-1}$   & $-0.02\pm0.01$ \\
\swift\        & PG~1247$+$267     & 21.7(8)    & $5.5 \times 10^{-3}$   & 0.14$\pm$0.09 \\
                  & PG~1634$+$706     & 35.7(18)  & $7.8 \times 10^{-3}$   &  0.07$\pm$0.02 \\
                  & HS~1700$+$6416   & 34.8(19) & $1.5 \times 10^{-2}$   & 0.14$\pm$0.04
\enddata
\tablenotetext{a}{The probability $p$ of the $\chi^2$ distribution, given the $\chi^2$ value and the degrees of freedom (dof).}
\label{tab:variability}
\end{deluxetable}

\subsubsection{Ground-Based Photometry}
\label{sec:ground_phot}

Optical imaging of the \chandra\ sources was performed with the 1~m telescope at the Tel Aviv University Wise Observatory (WO). Images were obtained with the PI CCD camera which has a $13\arcmin \times 13\arcmin$ field of view with a scale of $0.58''$~pix$^{-1}$. The sources were observed with a combination of several of the $g', r', i', z'$ filters (see Fukugita \et 1996) and Johnson $B, V, R, I$ filters which were available on particular nights of observations. The images were reduced in the standard way using
{\sc iraf}\footnote{IRAF (Image Reduction and Analysis Facility) is distributed by the National Optical Astronomy Observatories, which are operated by AURA, Inc, under cooperative agreement with the National Science Foundation.} routines. Broad-band light curves for the quasars were produced by comparing their instrumental magnitudes to those of non-variable stars in the field (see, e.g., Netzer \et 1996, for more details). The quoted uncertainties on the photometric measurements include the fluctuations due to photon statistics and the scatter in the measurement of the non-variable stars used. Flux calibration was done using the fluxes of a few tens of field stars from known catalogs. For PSS~0926$+$3055 and PSS~1326$+$0743 we used the SDSS catalog in order to flux calibrate the images in the $g', r', i', z'$ filters. To derive the flux for the field stars in the
$V, R, I$ bands we used the transformations between SDSS magnitudes and $V, R, I$ magnitudes as given by
R.~Lupton (2005).\footnote{http://www.sdss.org/dr5/algorithms/sdssUBVRITransform.html} For Q~0000$-$263 and BR~0351$-$1034 we used the USNO-A2.0 catalog and the prescription given by B.~Gary.\footnote{http://brucegary.net/dummies/USNO-A2\_Method.htm} We derived the
$B, V, R$ magnitudes of the field stars and used the transformations between SDSS and $B, V, R$ magnitudes of Jester \et (2005) to derive the flux in the $g'$ and $r'$ filters. The flux calibration performed here is only an approximation, due to the calibration prescriptions used, and there might be systematic uncertainties in the flux calibration on the order of $\sim0.5$~mag. These uncertainties are not included in the uncertainties quoted on the measurements below since we are interested only in the relative flux changes between the observed epochs.

\begin{figure}
\plotone{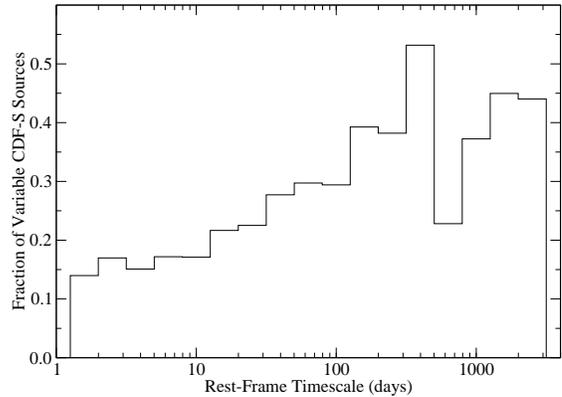}
\caption{Distribution of rest-frame timescales probed in the 2~Ms exposure of the \hbox{CDF-S}. These timescales are similar to those probed for our \chandra\ and \swift\ sources.}
\label{fig:timescales_2Ms}
\end{figure}

We also performed optical imaging of BR~0351$-$1034 with the du Pont 2.5~m telescope at Las Campanas Observatory (LCO). Images were obtained in the Johnson $B, V, R$ bands with the Wide Field CCD camera which has a scale of $0.484''$~pix$^{-1}$ and is equipped with a WF4K detector.\footnote{http://www.lco.cl/draft/direct-ccd-users-manual} The images were reduced and calibrated in the same manner as described above for the WO data. The observation log from the WO and LCO as well as the calibrated magnitudes are listed in Table~\ref{tab:ground}.

The photometric data of the \chandra\ sources were converted to flux densities based on the zero-point fluxes given in Fukugita \et (1996) for the
$g', r', i', z'$ filters, and in Bessell \et (1998) for the $B, V, R, I$ bands. The bands which are least contaminated by emission lines and that have the closest effective wavelengths to rest-frame 1450~\AA\ were used for obtaining $F_{\lambda}$(1450~\AA) values for each epoch. Given the fact that we use broadband filters with typical bandwidths of $\sim1000$~\AA\ and that the strongest emission lines in the rest-frame
\hbox{$\sim1200-1800$~\AA} accessible to us have typical equivalent widths (EWs) of $\ltsim 350$~\AA\ (given the redshifts of our sources), the maximum flux contributions of each of these lines to each of our filters should be $\ltsim35$\% (see also, e.g., Elvis \et 2012). For example, one of the strongest emission lines in the observed-frame optical band, C~{\sc iv}~$\lambda1549$, has EW$\sim100$~\AA\ in PSS~0926$+$3055 and PSS~1326$+$0743 (obtained from our ground-based spectroscopy; see below), as well as in Q~0000$-$263 (see Schneider \et 1989), hence the contribution of this emission line to the flux densities obtained in, e.g., the $i'$-band is of the order of $\sim10$\%, not considerably larger than the typical photometric uncertainties (see Table~\ref{tab:ground}; there are no published values of emission-line EWs for BR~0351$-$1034).

\begin{figure*}
\plotone{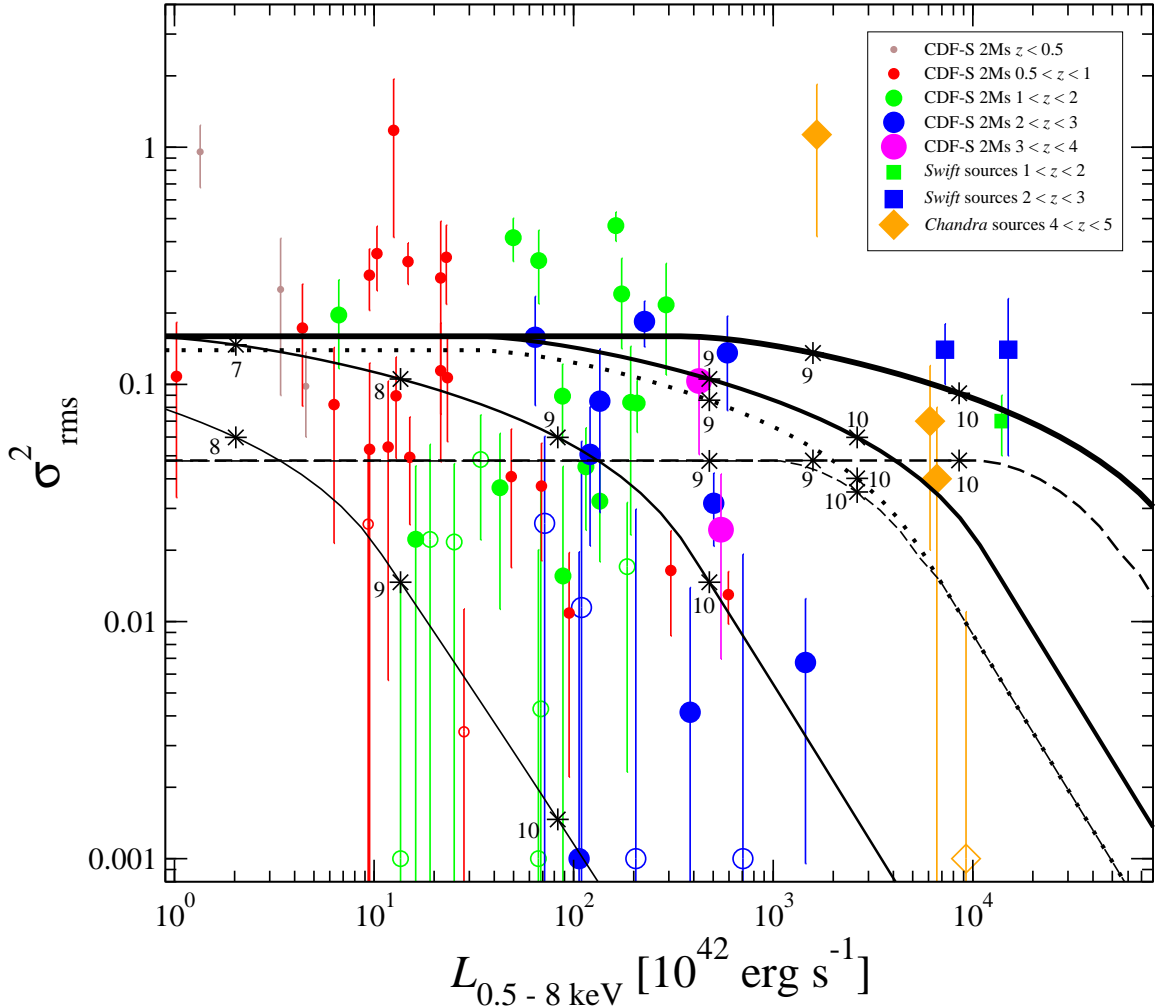}
\caption{Excess variance versus luminosity in the observed-frame $0.5-8$~keV band. Circles represent \xray-selected AGNs from the 2~Ms exposure of the CDF-S survey. Squares and diamonds represent our \swift\ and \chandra\ sources, respectively. Symbol sizes increase with increasing source redshift, and open symbols indicate non-variable sources (brown, red, green, blue, magenta, and orange symbols mark sources at $z<0.5$, $0.5<z<1$, $1<z<2$, $2<z<3$, $3<z<4$, $4<z<5$, respectively, in the online version). Sources with negative $\sigma^2_{\rm rms}$ values
have been assigned a fixed excess variance of $\sigma^2_{\rm rms} = 0.001$. Error bars on $\sigma^2_{\rm rms}$ represent `formal' errors due only to flux measurement errors and not those due to red-noise intrinsic scatter. Simulations of $\sigma^2_{\rm rms}$ as a function of \xray\ luminosity, using the Papadakis \et (2008) model, are represented by: 1) continuous lines using $T_{\rm max, rest} = 8$~yr, $T_{\rm min, rest} = 1$~d, and
\lledd$ = 0.001, 0.01, 0.1, 0.5$, where line thickness increases with increasing \lledd, 2) dashed lines using $T_{\rm max, rest} = 3$~yr,
$T_{\rm min, rest} = 100$~d, and \lledd$ = 0.1, 0.5$, where line thickness increases with increasing \lledd, and 3) a dotted line using
$T_{\rm max, rest} = 3$~yr, $T_{\rm min, rest} = 1$~d, and \lledd$ = 0.1$. Stars with annotated values along each curve indicate $\log$~\mbh\ values (in units of \msun) derived from the corresponding \xray\ luminosity and the assumed \lledd.}
\label{fig:Paolillo04_fig12}
\end{figure*}

The optical fluxes and the corresponding \aox\ values are given in Table~\ref{tab:lc_chandra_opt} where we also list the time separations between the optical and \chandra\ observations that were used for computing \aox. These time separations are on the order of $\approx1$~d in the rest frame. Based on the photometry in Table~\ref{tab:ground} and on the results of the \swift\ sources (\S~\ref{sec:uvot}), we do not consider these time delays to be significant as we do not detect large optical flux variations on such short timescales. However, we do detect optical flux variations at a level of up to $\sim20$\% on significantly longer timescales. Therefore, similar to the \swift\ sources, we find that \aox\ variability is dominated by the \xray\ variability, except for PSS~1326$+$0743 which is not \xray\ variable.

\subsubsection{Optical Spectroscopy}
\label{sec:opt_spec}

We searched for optical spectroscopic variations in two of our \chandra\ sources, which we observed within $\sim1$~d in the rest frame from their \chandra\ Cycle~12 observations, by comparing their spectra with archival data. Spectroscopic observations of PSS~0926$+$3055 and
PSS~1326$+$0743 were performed with the queue-scheduled 9~m {\sl Hobby-Eberly Telescope} ({\sl HET}; Ramsey \et 1998; Shetrone \et 2007)
on 2011~March~1 and 2011~March~10, respectively. Observations of both sources were obtained with the Marcario Low-Resolution Spectrograph (Hill \et 1998) using the G1 grism with the $3''$ slit, providing a resolution of $R\sim240$, and the OG515 blocking filter. The spectrum of each source was obtained in two sub-exposures of 450~s each in order to remove cosmic rays. Image reduction and spectral analysis were performed with
{\sc iraf} using neon and argon wavelength calibration lamps. Flux calibration was performed using a spectrophotometric standard star taken each night and an $I$-band image snapshot that could be compared to SDSS $i$-band magnitudes of nearby field stars. The final, calibrated spectra are shown in Fig.~\ref{fig:HET_spectra}.

The flux density at rest-frame 1450~\AA\ extracted from our spectrum of PSS~0926$+$3055 is a factor of $\sim2$ larger than the value derived from
the SDSS $i$-band magnitude of the source, \hbox{$i=17.07$}, which is only 0.06 mag fainter than the $i'$-band magnitude we measure on 2011 March 4, as close as possible to the time of our {\sl HET} observation. The {\sl HET} flux density we measure for this source is also $\sim60$\% larger than the one obtained by Vignali \et (2003) who used {\sl HET} spectroscopy with a different standard-star calibration. For PSS~1326$+$0743, our {\sl HET} measurement gives a flux density at rest-frame 1450\,\AA\ which is only $\sim5$\% larger than the value obtained from its SDSS
\hbox{$i$-band} magnitude, \hbox{$i=17.50$}, and is consistent with our $i'$-band measurement from 2011 March 14, as close as possible to our
{\sl HET} observation; the value obtained by Vignali \et (2003) using {\sl HET} spectroscopy of this source is a factor of $\sim2$ lower. The large differences in flux density indicate uncertainties on the order of $\approx0.5-1$~mag in the flux calibration between the {\sl HET} spectra and the photometry as well as between the two pairs of {\sl HET} spectra. These large uncertainties likely originate from observing a different calibration star at a different position on the sky with respect to the telescope in each epoch as well as from slit losses due to the fact that the {\sl HET} spectra were not necessarily obtained during photometric conditions. By fitting a power-law continuum and two Gaussians to model the spectral region around the C~{\sc iv} emission line, we find that the C~{\sc iv} EW in both sources agrees to within $\sim5$\% with the values given in Vignali \et (2003).

\begin{figure}
\plotone{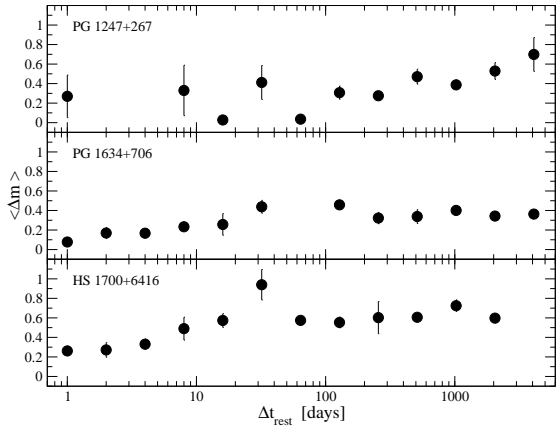}
\caption{Mean SFs of the \swift\ sources. Average magnitude difference in each time bin is plotted as a function of rest-frame time interval. These sources are more variable on longer timescales than on shorter ones, and significant \xray\ variability is detected on rest-frame timescales as short as $\sim1$~d.}
\label{fig:SF_Swift}
\end{figure}

\subsection{\xray\ spectral Variability}
\label{sec:spectral}

By design, our economical \xray\ observations are intended to provide only the minimal number of counts that are sufficient for basic time-series analyses of intensity fluctuations. Naturally, this approach does not allow meaningful \xray\ spectral measurements. Table~\ref{tab:chandra_counts} shows that the effective power-law photon indices of our \chandra\ sources are consistent between \chandra\ Cycles~12 and 13 and with measurements obtained from high-quality \xray\ spectroscopy (see Shemmer \et 2005). Among the \swift\ sources, only PG~1247$+$267 and PG~1634$+$706 have sufficient numbers of counts ($\sim65-350$) in three and seven \swift\ observations, respectively, to perform basic \xray\ spectral analysis (see Table~\ref{tab:swift_log}). We analyzed the spectrum extracted from each of these {\sl XRT} observations using
{\sc xspec}~v12 (Arnaud 1996) with a Galactic absorbed power-law model in the observed-frame $0.2-10$~keV band. The resultant photon indices were consistent for each source within their errors (at 90\% confidence) among the different epochs. In particular, we do not detect any evidence for steepening in the \xray\ spectra of PG~1247$+$267 and PG~1634$+$706 as they become \xray\ brighter by factors of $\sim2$ and $\sim4$, respectively (cf. Mateos \et 2007; Sobolewska \& Papadakis 2009; Gibson \& Brandt 2012). We also analyzed the \chandra\ spectra of
PG~1634$+$706 in the same manner, using the {\sc ciao} task {\sc psextract}, and found that the photon index of the source in the observed-frame $0.5-8$~keV band did not vary significantly (at 90\% confidence) between the different epochs (although most of these are within a few hours in the rest frame from each other).

HS~1700$+$6416 has been known to display complex \xray\ spectral variability, although the power-law photon index remains constant within the errors (e.g., Lanzuisi \et 2012). The spectral variability is associated primarily with variations in the intrinsic absorption column density and with
potential variable relativistic outflows. Our analysis of the \chandra\ data of the source is consistent with the Lanzuisi \et (2012) findings. Unfortunately, the low count rate of the {\sl XRT} observations prevents us from obtaining additional insight into the nature of these spectral variations.

\begin{figure}
\plotone{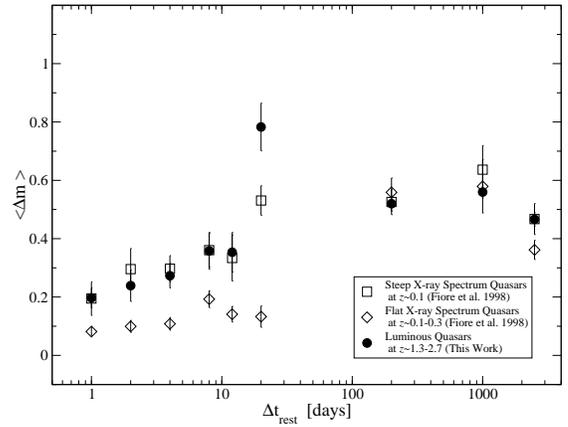}
\caption{Ensemble SF of the \swift\ sources (circles) compared to the ensemble SFs of the steep- and flat-\xray-spectrum quasars at low redshift, marked by squares and diamonds, respectively, adapted from Fiore \et (1998). Units are similar to those in Fig.~\ref{fig:SF_Swift}. The temporal behavior of the \swift\ sources is remarkably similar to that of steep-\xray-spectrum quasars, perhaps due to the fact that both groups have similarly high accretion rates.}
\label{fig:SF_compare}
\end{figure}

\section{DISCUSSION}
\label{sec:discussion}

The basic finding from our exploratory, long-term \xray\ monitoring program is that significant \xray\ flux variations persist at the highest AGN luminosities and redshifts. In this section, we discuss potential sources for the persistent variability and whether any of these are expected to evolve with cosmic time. We stress again that due to the nature of this pilot project and the small sample size the following discussion is mostly qualitative; a more rigorous treatment is beyond the scope of this work.

\begin{deluxetable*}{lllcccccccc}
\tablecolumns{11}
\tablecaption{Optical Photometry}
\tablehead{ 
\colhead{} &
\colhead{} &
\colhead{Obs.} &
\colhead{$g'$} &
\colhead{$r'$} &
\colhead{$i'$} &
\colhead{$z'$} &
\colhead{$B$} &
\colhead{$V$} &
\colhead{$R$} &
\colhead{$I$} \\
\colhead{Quasar} &
\colhead{Obs.} &
\colhead{Date} &
\colhead{(mag)} &
\colhead{(mag)} &
\colhead{(mag)} &
\colhead{(mag)} &
\colhead{(mag)} &
\colhead{(mag)} &
\colhead{(mag)} &
\colhead{(mag)}
}
\startdata
Q~0000$-$263
& WO  & 2011 Sep 4  & 18.93$\pm$0.02 & 17.45$\pm$0.02 &        \nodata &        \nodata & 
                      19.58$\pm$0.04 & 18.23$\pm$0.02 & 17.16$\pm$0.02 &        \nodata    \\
& WO  & 2012 Sep 14 & 18.93$\pm$0.03 & 17.48$\pm$0.01 &        \nodata &        \nodata & 
                      19.45$\pm$0.09 & 18.28$\pm$0.02 & 17.18$\pm$0.03 &        \nodata    \\
& WO  & 2012 Sep 15 & 18.97$\pm$0.02 & 17.48$\pm$0.01 &        \nodata &        \nodata & 
                      19.53$\pm$0.04 & 18.26$\pm$0.02 & 17.17$\pm$0.01 &        \nodata    \\
BR~0351$-$1034	
& WO  & 2011 Mar 3  &        \nodata & 19.39$\pm$0.06 &        \nodata &        \nodata & 
                             \nodata &        \nodata & 19.24$\pm$0.05 &        \nodata    \\
& WO  & 2011 Mar 5  &        \nodata & 19.33$\pm$0.04 &        \nodata &        \nodata & 
                             \nodata &        \nodata &        \nodata &        \nodata    \\
& WO  & 2011 Sep 26 &        \nodata & 19.33$\pm$0.03 &        \nodata &        \nodata & 
                             \nodata & 20.59$\pm$0.09 & 19.29$\pm$0.04 &        \nodata    \\
& LCO & 2011 Oct 29 &        \nodata &        \nodata &        \nodata &        \nodata & 
                      22.79$\pm$0.11 & 20.55$\pm$0.02 & 19.35$\pm$0.03 &        \nodata    \\
PSS~0926$+$3055 
& WO  & 2011 Mar 4  & 18.45$\pm$0.01 & 17.13$\pm$0.01 & 17.01$\pm$0.01 & 17.22$\pm$0.03 & 
                             \nodata & 17.83$\pm$0.02 & 16.90$\pm$0.01 & 16.60$\pm$0.02   \\
& WO  & 2012 Feb 4  & 18.55$\pm$0.05 & 17.35$\pm$0.11 & 17.19$\pm$0.06 &        \nodata & 
                             \nodata & 17.94$\pm$0.05 & 17.11$\pm$0.08 & 16.66$\pm$0.04   \\
PSS~1326$+$0743
& WO  & 2011 Mar 8  & 19.15$\pm$0.10 &        \nodata &        \nodata &        \nodata & 
                             \nodata & 18.47$\pm$0.03 & 17.48$\pm$0.02 & 16.88$\pm$0.03   \\
& WO  & 2011 Mar 14 & 19.28$\pm$0.03 & 17.82$\pm$0.10 & 17.51$\pm$0.10 & 17.15$\pm$0.03 & 
                             \nodata & 18.47$\pm$0.02 & 17.49$\pm$0.02 & 16.77$\pm$0.12   \\
& WO  & 2012 May 1  &        \nodata & 17.79$\pm$0.06 & 17.61$\pm$0.07 &        \nodata & 
                             \nodata & 18.52$\pm$0.14 & 17.59$\pm$0.07 & 16.69$\pm$0.09       
\enddata
\label{tab:ground}
\end{deluxetable*}

\subsection{Were Quasars More X-ray Variable in the Early Universe?}
\label{sec:evolution}

Six of our seven luminous RQQs display pronounced \xray\ variability similar to many of their nearby and far less-luminous counterparts
(e.g., Leighly 1999; Turner \et 1999, Markowitz \& Edelson 2004; Ponti \et 2012; Vagnetti \et 2013). The rest-frame energy band over which the \xray\ variability is measured, $\sim2.6-10.4$~keV, is also comparable to the rest-frame energy bands used in surveys of lower-redshift sources. While their variability amplitudes do not exceed those of their lower-luminosity counterparts, our sources do not follow the well-known variability-luminosity
anticorrelation observed in the nearby universe (e.g., Lawrence \& Papadakis 1993; P04). This finding is at odds with simplistic light-crossing time considerations since the most luminous sources are expected to host the largest continuum emission regions and thus their light fluctuations are expected to be suppressed.

P04 have found that $\sigma^2_{\rm rms}$ decreases with \xray\ luminosity in the observed-frame $0.5-7$~keV band ($L_{\rm 0.5-7}$)
up to $L_{\rm 0.5-7} \sim10^{44}$~erg~s$^{-1}$, and increases slightly beyond that luminosity (see their Fig.~12), a result supported by the CDF-S 2~Ms data shown here (Fig.~\ref{fig:Paolillo04_fig12}). This `upturn' in the amplitude-luminosity relation could also be interpreted as an amplitude-redshift relation and an increase in the fraction of variable sources as a function of redshift beyond $z\sim2$ due to the flux-limited nature of the
CDF-S survey. Similar trends have been reported previously by Almaini \et (2000) and Manners \et (2002). Vagnetti \et (2011) report a stronger dependence of \xray\ variability on luminosity and a much weaker one on redshift. However, Mateos \et (2007) do not find a significant dependence between \xray\ variability and either luminosity or redshift in the \xmm\ Lockman Hole survey, and Gibson \& Brandt (2012) do not find an increase of \xray\ variability with redshift up to $z\sim4$. Although we do not detect an increase in $\sigma^2_{\rm rms}$ with respect to either $L_{\rm 0.5-8}$ or redshift beyond the limits of the 2~Ms CDF-S survey, our variable sources have an average variability amplitude that is similar to the bulk of
lower-luminosity sources at $z\ltsim1$ (Fig.~\ref{fig:Paolillo04_fig12}). This trend is difficult to explain if high-redshift sources follow the same variability-luminosity relation of their lower-redshift counterparts, which is driven primarily by BH mass (see \S~\ref{sec:introduction}). Furthermore, the median of the $\sigma^2_{\rm rms}$ values of our \swift\ sources, at $\left < z \right >\simeq2.0$, is larger than the median of the
$\sigma^2_{\rm rms}$ values of our variable \chandra\ sources at $\left < z \right >\simeq4.2$. Notwithstanding the large differences in the sampling patterns between the \chandra\ and \swift\ sources, as well as the uncertainties associated with computing $\sigma^2_{\rm rms}$ for individual sources (see \S~\ref{sec:results_amplitudes}), this is an indication that, at least within the limits of our sample, the \xray\ variability amplitude does not depend primarily on redshift, and therefore the dependence of \xray\ variability on additional parameters should be explored.

\subsection{Black-Hole Mass and Accretion Rate Effects}
\label{sec:accretion}

Since AGN luminosity is not the only parameter controlling the variability amplitude, we must consider the combined effects of the monitoring duration, and the source BH mass (\mbh) and accretion rate on our results (e.g., McHardy \et 2006). Although we detect significant \xray\ variability on all rest-frame timescales $>1$~d at least for the \swift\ sources, we find that the variability amplitude increases with the temporal baseline up to rest-frame timescales of $\approx10^3$~d. This is in agreement with numerous studies of AGN variability, in the \hbox{X-rays} and in other bands, and is consistent with the generic AGN PSD function where more power is observed at longer timescales (e.g., Fiore \et 1998; Cid-Fernandes \et 2000; Uttley \et 2002; Markowitz \& Edelson 2004; Mushotzky \et 2011). Fig.~\ref{fig:SF_compare} indicates that, within the limits of our sample, the \xray\ variability behavior of luminous high-redshift sources is not significantly different from that of nearby AGNs, in terms of their SF (see also Vagnetti \et 2011). Nevertheless, continued \xray\ monitoring is required, especially for the \chandra\ sources, in order to perform a more detailed comparison between nearby and distant sources. Ideally, what future \xray\ missions should provide is a dense and regularly sampled \xray\ light curve of luminous, high-redshift sources from which high-quality PSD functions can be computed.

Given the fact that AGN variability increases on longer timescales, AGN variability studies based on survey data are expected to show smaller variability amplitudes for higher-redshift sources due to cosmic time dilation. Our monitoring project is not subject to such bias since our temporal baseline is not fixed and varies from source to source. The rest-frame temporal baselines for our sources are in the range $\approx800-3000$~d,
comparable to the temporal baselines for the bulk of the sources in the 2~Ms exposure of the CDF-S (Fig.~\ref{fig:timescales_2Ms}). Given the negative power-law slope of the PSD function, the variability amplitudes of our sources (and most of the \hbox{CDF-S} sources) are expected to increase as their temporal baselines increase. Our monitoring project should therefore continue in order to allow a much better comparison with the CDF-S sources by utilizing the planned 7~Ms exposure of the CDF-S, thus extending the rest-frame temporal baselines to $\approx1000-5000$~d.
The continued monitoring will also enable searching for the expected flattening in the PSD function of luminous AGNs.

Since luminosity is a rough proxy to \mbh, the amplitude-luminosity anticorrelation may be based on an amplitude-\mbh\ anticorrelation which is due to light-crossing time effects. As \mbh\ increases, the PSD function shifts to lower frequencies, resulting in a decrease in variability amplitude (which is equivalent to integrating the PSD function between two specific frequencies). In addition, the variability amplitude may increase when the normalized accretion rate (in terms of \lledd) increases (e.g., McHardy \et 2006, but see also, e.g., Gonz{\'a}lez-Mart{\'{\i}}n \& Vaughan 2012;
Ponti \et 2012). Fig.~\ref{fig:Paolillo04_fig12} may therefore include a combination of different amplitude-luminosity correlations, depending on different values of \lledd\ or on different accretion modes at different luminosities or redshifts. These correlations may have different slopes and constants  in the $\sigma^2_{\rm rms}$-$L_{\rm 0.5-8}$ plane and thus contribute to the upturn and the subsequent flattening in that diagram (Allevato \et 2010). Testing this possibility requires monitoring of faint AGNs at high redshift ($z\gtsim4$) which is extremely challenging. Such low-\mbh\ sources may vary even more than the luminous sources in our sample.

According to Papadakis \et (2008), the upturn and then the flattening observed in the amplitude-luminosity relation in \xray\ survey data is due to
the fact that as the redshift increases, luminosity increases as well (due to selection effects), but more rapidly than the increase in \mbh, resulting in an increase in \lledd; the increase in \lledd\ contributes to the increase in the variability amplitude. As the redshift increases even further, the increase in amplitude must stop due to the effect of cosmic time dilation since the temporal baseline is fixed. In our case, however, the temporal baseline increases continuously, and it is expected that the variability amplitude will continue to increase up to a timescale where the SF flattens. If the trend of having roughly constant $\sigma^2_{\rm rms}$ values as $L_{\rm 0.5-8}$ increases is due to a combination of \mbh\ and \lledd\ alone, it requires that the decrease in amplitude, due to the increase in \mbh, approximately cancels the increase in amplitude as \lledd\ increases (assuming the temporal baseline is sufficiently long).

We have simulated the combined effects of \mbh\ and \lledd\  on $\sigma^2_{\rm rms}$ using the Papadakis \et (2008) model, and plotted the results
for several \lledd\ values and three different rest-frame sampling patterns in Fig.~\ref{fig:Paolillo04_fig12}. In short, this model assumes a double power-law PSD function with a constant normalization,\footnote{Ponti \et (2012) suggest that the normalization of the PSD function depends on the accretion rate.} where the power law changes slope from $\alpha = -2$ at the highest frequencies to $\alpha = -1$ at frequencies below a break frequency, $\nu_{\rm break}$, where \hbox{$\nu_{\rm break} \propto \eta$(\lledd)\mbh$^{-1}$} and $\eta$ is the accretion efficiency, taken here as
$\eta = 0.1$ (following McHardy \et 2006). We have integrated the model PSD function in the frequency domain using 8~yr and 1~d for the lowest and highest rest-frame frequencies, respectively, and computed $\sigma^2_{\rm rms}$ as a function of $L_{\rm 0.5-8}$, using the Marconi \et (2004) relation between the \xray\ and bolometric luminosities. These frequencies are representative of the total duration and of the minimum sampling period of our \swift\ sources (considering the binned \swift\ data for these sources).

\begin{deluxetable}{llcccl}
\tablecolumns{6}
\tablecaption{Optical Light Curve and \aox\ Data for the \chandra\ Sources}
\tablehead{
\colhead{Quasar} &
\colhead{JD} &
\colhead{$F_{\lambda}$\tablenotemark{a}} &
\colhead{\aox} &
\colhead{$\Delta t$\tablenotemark{b}} &
\colhead{Obs.}
}
\startdata
Q~0000$-$263 & 2455809.5 & 2.41$\pm$0.04 & $-1.74$ & $1.2$ & WO \\
& 2456185.5 & 2.35$\pm$0.07 & $-1.76$ & $2.2$ & WO \\
& 2456186.5 & 2.39$\pm$0.03 & \nodata & \nodata & WO \\
BR~0351$-$1034 & 2455624.2  & 0.33$\pm$0.02 & \nodata & \nodata & WO \\
& 2455626.2  & 0.37$\pm$0.01 & \nodata & \nodata & WO \\
& 2455831.5  & 0.31$\pm$0.01 & $-1.65$ & $0.6$ & WO \\
& 2455864.8  & 0.30$\pm$0.01 & $-1.67$ & $0.4$ & LCO \\
PSS~0926$+$3055 & 2455625.2 & 2.99$\pm$0.03 & $-1.75$ & $0.3$ & WO \\
& 2455962.3 & 2.53$\pm$0.14 & $-1.78$ & $4.4$ & WO \\
PSS~1326$+$0743 & 2455629.6 & 1.75$\pm$0.03 & $-1.65$ & $0.4$ & WO \\
& 2455635.5 & 1.74$\pm$0.03 & \nodata & \nodata & WO \\
& 2456049.3 & 1.58$\pm$0.11 & $-1.64$ & $0.3$ & WO
\enddata
\tablecomments{For each source, \aox\ is given only for the shortest time separations between the optical and \chandra\ observations.}
\tablenotetext{a}{Flux density at rest-frame 1450~\AA\ in units of $10^{-16}$~erg~cm$^{-2}$~s$^{-1}$~\AA$^{-1}$.}
\tablenotetext{b}{Rest-frame days between the optical and \chandra\ observations.}
\label{tab:lc_chandra_opt}
\end{deluxetable}

Fig.~\ref{fig:Paolillo04_fig12} demonstrates that using a constant \lledd$ = 0.5$ for computing $\nu_{\rm break}$, produces a predicted
$\sigma^2_{\rm rms}$ consistent with those of our \swift\ sources; it also shows that using lower \lledd\ values result in lower predicted
$\sigma^2_{\rm rms}$ values. In fact, \mbh\ values derived using the single-epoch spectroscopic method based on the \hb\ line for two of our \swift\ sources, PG~1247$+$267 and PG~1634$+$706, are \hbox{$1.4 \times 10^{10}$}~\msun\ and \hbox{$3.9 \times 10^{10}$}~\msun, respectively (with a typical uncertainty of a factor of $\sim2-3$ on the derived mass), and their respective \lledd\ values are 0.5 and 0.3 (see Shemmer \et 2006b and references therein for more details). These values are consistent with our model \lledd\ and the derived \mbh. Additionally, the McHardy \et (2006) relation for the break frequency, using the \hb-based \mbh\ and \lledd\ values of PG~1247$+$267 and PG~1634$+$706, gives
\hbox{$\nu_{\rm break} = 1.04 \times 10^{-7}$}~Hz and  \hbox{$\nu_{\rm break} = 2.23 \times 10^{-8}$}~Hz, respectively, which are well within the temporal windows of both of these sources. While the \hb-based \mbh\ values are among the highest measured for any AGN, the \lledd\ values of the two \swift\ sources are typical of quasars and somewhat lower than those of nearby narrow-line Seyfert~1 galaxies. Indeed, this model is also consistent with the $\sigma^2_{\rm rms}$ values of several low-redshift AGNs from the CDF-S survey, which have a roughly similar sampling pattern; such sources, as well as those that lie above this model in Fig.~\ref{fig:Paolillo04_fig12}, may have relatively high accretion rates (although the reliability of single-source measurements must be interpreted with caution; see Allevato \et 2013).

\begin{figure}
\epsscale{1.1}
\plotone{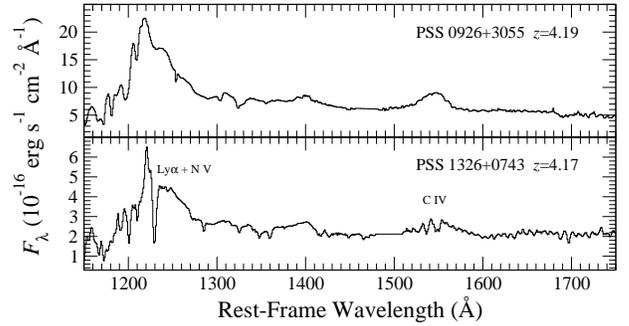}
\caption{New {\sl HET} spectra of PSS~0926$+$3055 ({\it top}) and PSS~1326$+$0743 ({\it bottom}). Prominent emission lines are indicated in the bottom panel. The emission line EWs in these sources have not significantly changed compared to the respective archival {\sl HET} spectra of Vignali \et (2003).}
\label{fig:HET_spectra}
\end{figure}

We ran three additional models of this kind. The first two were designed to match the temporal sampling pattern of the \chandra\ sources, using 3~yr and 100~d for the lowest and highest rest-frame frequencies, respectively; one of these uses \lledd$ = 0.1$ and the other uses \lledd$ = 0.5$.
Fig.~\ref{fig:Paolillo04_fig12} shows that the model using \lledd$ = 0.5$ predicts $\sigma^2_{\rm rms}$ values that are consistent with those of two of the variable \chandra\ sources, Q~0000$-$263 and PSS~0926$+$3055. The third model uses 3~yr and 1~d for the lowest and highest rest-frame frequencies, respectively, and \lledd$ = 0.1$, mimicking the temporal sampling pattern of a typical variable AGN from the 2~Ms exposure of the
CDF-S at $z\sim1.5$, which is the average redshift of the sources in this survey. This model also appears to predict $\sigma^2_{\rm rms}$ values similar to those observed for several CDF-S sources with assumed \mbh\ values of $<10^9$~\msun. At least to first order, it appears as if the combined effects of \mbh\ and \lledd\ can largely explain the observed \xray\ variability amplitudes of AGNs over the widest possible ranges of \xray\ luminosity and redshift.

\subsection{Additional Contributions to \xray\ Variability in RQQs?}
\label{sec:jets}

The source of \xray\ variability in AGNs is likely much more complex than the combined effects of \mbh\ and accretion power on a single source PSD function. For example, spectroscopic observations show that variable obscuration close to the central engine plays a role in the \xray\ variability of AGNs (e.g., Risaliti \et 2009); however, our sources are not expected to be subject to significant obscuration due to their extremely high luminosities (e.g., Hasinger 2008). Except for the \xray\ spectral variations detected in HS~1700$+$6416 (Lanzuisi \et 2012), we do not detect any major spectral changes in our sources due to the lack of higher-quality multi-epoch spectroscopic observations. Another contribution to \xray\ variability may be due to a more distant reflector which can dilute the fluctuations of the more compact and highly variable nuclear \xray\ source. While luminous quasars are not expected to exhibit a strong Compton reflection component in their \xray\ spectra, due to the diminishing covering factor of a putative torus of gas and dust as AGN luminosity increases (e.g., Ueda \et 2003; Maiolino \et 2007; Ricci \et 2013), two of our sources, PG~1247$+$267 and PG~1634$+$706, actually show hints of Compton reflection components (Shemmer \et 2008 and references therein). Hence it is not clear if and to what extent an \xray\ reflector can play a role in the \xray\ variability of luminous quasars.

The fact that at least two of our sources, PG~1634$+$706 and HS~1700$+$6416, exhibit significant variability on timescales as short as a few days in the rest frame is consistent with the idea that the main contribution to the \xray\ variability arises from a small region that may not scale with, or have only a weak dependence on, luminosity (or \mbh). Since $\nu_{\rm break}$ of the \xray\ PSD function is expected to decrease as \mbh\ increases, we would have expected much slower variations in our sources with respect to nearby AGNs. However, we do not detect such a trend in our SFs. In particular, based on Fig.~\ref{fig:SF_compare}, we do not observe the strong dependence of the SF on luminosity as reported in Vagnetti \et (2011; cf. their Fig.~10), who find that the slope of the SF increases with increasing luminosity; a dependence that is not observed for the PSD function. As we explain above, high \lledd\ values may compensate for the reduced variability due to the increase in \mbh\ and contribute to more rapid variability as higher frequencies receive more power.

In addition to the effects of the accretion rate, variable jet emission may contribute to the relatively rapid and significant \xray\ variability we observe in our luminous sources. Although our selection of RQQs was intended to minimize such `contamination', RQQs are not necessarily `radio silent' (e.g., Kellermann \et 1994; Barvainis \et 1996; Falcke \et 1996; Blundell \& Beasley 1998; Blundell \et 2003). However, we do not expect that the relatively weak \xray\ emission from the jets in RQQs can dominate the observed \xray\ variability (e.g., Miller \et 2011). In order to quantify, or constrain, the contribution of the jet to \xray\ variability in RQQs, and since the radio emission in RQQs may be largely due to star formation (e.g., Padovani \et 2011), future monitoring campaigns should perform near simultaneous \xray\ and radio observations of such sources across a wide luminosity range (e.g., Barvainis \et 2005).

\section{SUMMARY}
\label{sec:conclusions}

We present initial results from \chandra\ and \swift\ monitoring of a sample of seven luminous, RQQs at $1.33\leq z\leq 4.35$ extending over
$\approx800-3000$~d in the rest frame. Our \xray\ observations are supported by archival \xray\ data and by simultaneous or nearly simultaneous optical-UV observations, allowing a qualitative investigation of quasar variability properties at the highest luminosities and redshifts. Our main findings can be summarized as follows:

\begin{enumerate}

\item RQQs exhibit an excess in \xray\ variability, above the well-known amplitude-luminosity anticorrelation, almost independent of luminosity or redshift, at $L\gtsim10^{44}$~erg~s$^{-1}$ in the \hbox{0.5--8}~keV band. We suggest that this excess is primarily due to the higher accretion rates at higher luminosities and redshifts.

\item We find no direct evidence of evolution in the \xray\ variability of RQQs up to \hbox{$z\sim4.2$}, but a firmer conclusion will require continued monitoring, especially of our \chandra\ sources.

\item Our sources vary more on longer timescales than on shorter timescales, in agreement with AGN behavior at lower luminosities or redshifts, and some exhibit significant variability on timescales as short as $\sim1$~d in the rest frame.

\item We find significant variations in the \aox\ values of our \swift\ sources, implying relative flux variations at a level up to a factor of $\sim3$, which are dominated by the \xray\ variations. We confirm earlier reports that the dispersion in the luminosity-corrected \aox\ distribution of type~1 RQQs may be dominated by variability.

\item Except for HS~1700$+$6416, we do not detect significant \xray\ spectral variations in our sources, even for corresponding flux variations of up to a factor of $\sim4$.

\end{enumerate}

We plan to continue this ongoing study using \chandra\ in order to characterize better the \xray\ variability properties of our four sources at
\hbox{$4.10\leq z\leq 4.35$} that currently have only four sparsely sampled \xray\ epochs. In particular, our aim is to quantify their variability timescales by means of SFs. We also plan to characterize better the temporal behavior of our three sources at $1.33\leq z \leq 2.74$ using \swift.
Our results will serve as a benchmark for future \xray\ variability studies of distant AGNs, e.g., with the planned CDF-S 7~Ms survey, or with
{\sl eROSITA} and its multiepoch, all-sky survey spanning $\gtsim4$~yr.

\acknowledgements

The scientific results reported in this article are based on observations made by the Chandra \xray\ Observatory and on data obtained from the Chandra Data Archive. Support for this work was provided by the National Aeronautics and Space Administration through Chandra Award Numbers GO1-12132X and GO2-13120X (O.~S) issued by the Chandra \xray\ Observatory Center, which is operated by the Smithsonian Astrophysical Observatory for and on behalf of the National Aeronautics and Space Administration under contract NAS8-03060. We also gratefully acknowledge support from NASA \swift\ grant \hbox{NNX08AT26G} (O.~S), NASA ADP grant \hbox{NNX10AC99G} (W.~N.~B), the Kitzman Fellowship at the Technion (S.~K), and Fondecyt Project \#1120328 (P.~L). This work is based, in part, on observations obtained with the Tel Aviv University Wise Observatory 1~m telescope. The {\sl HET} is a joint project of the University of Texas at Austin, the Pennsylvania State University, Stanford University,
Ludwig-Maximillians-Universit\"{a}t M\"{u}nchen, and Georg-August-Universit\"{a}t G\"{o}ttingen. The {\sl HET} is named in honor of its principal benefactors, William P. Hobby and Robert E. Eberly. We thank Doron Chelouche and Dipankar Maitra for fruitful discussions. This research has made use of the NASA/IPAC Extragalactic Database (NED) which is operated by the Jet Propulsion Laboratory, California Institute of Technology, under contract with the National Aeronautics and Space Administration. This research has also made use of data provided by the High Energy Astrophysics Science Archive Research Center (HEASARC), which is a service of the Astrophysics Science Division at NASA/GSFC and the High Energy Astrophysics Division of the Smithsonian Astrophysical Observatory.

{{\it Facilities:} \facility{CXO (ACIS)}, \facility{Swift (XRT, UVOT)}}

\end{document}